\documentclass[runningheads]{llncs}

\newif\ifappendix
\appendixtrue
\ifappendix
\else
\def\marginpar#1{}
\fi

\usepackage{amsmath,amssymb}
\usepackage{mathtools}  % \mathrlap...
\usepackage[vlined]{algorithm2e}
\usepackage[hypertexnames=false]{hyperref}
\usepackage{standalone}
\usepackage{enumitem}
\usepackage{booktabs}
\usepackage{longtable}
\usepackage{xcolor}
\usepackage{colortbl}
\usepackage[disable]{todonotes}
\usepackage{fontawesome5}
\usepackage{adjustbox}
\usepackage{xspace}

\usepackage{tikz}
\usetikzlibrary{automata}
\usetikzlibrary{arrows.meta}
\usetikzlibrary{bending}
\usetikzlibrary{calc}
\usetikzlibrary{positioning}
\usetikzlibrary{shapes.geometric}
\usetikzlibrary{shapes.symbols}
\tikzset{
      >={Stealth[round,bend]},
      StealthDouble/.tip={Stealth[round,quick,length=5.75pt]},
      double>/.style={-StealthDouble,double},
      root/.style={draw,rectangle,dashed},
      inode/.style={draw,circle,inner sep=2pt},
      low/.style={->,densely dotted},
      high/.style={->},
      termn/.style={draw,rectangle},
      terma/.style={draw,rectangle,double},
      rlink/.style={->,dashed},
      win/.style={fill=green!20},
      lose/.style={fill=red!20},
      control/.style={draw,inner sep=1pt,diamond,aspect=1},
      automaton/.style={
        shorten >=1pt,>={Stealth[round,bend]},
        node distance=2cm,
        initial text=,
        every initial by arrow/.style={every node/.style={inner sep=0pt}},
      every state/.style={
        align=center,
        fill=white,
        minimum size=7.5mm,
        inner sep=0pt,
        execute at begin node=\strut,
      },%
   }
}

\usepackage{mathrsfs}           % For \mathscr
\SetProcNameSty{texttt}
\SetKwProg{Fn}{Function}{}{end}
\SetVlineSkip{0.0em}
\SetKwComment{tcp}{// }{}
\SetKwInOut{Input}{input}
\SetKwInOut{Output}{output}
\SetKwInput{Var}{var}
\SetKwFunction{mk}{makebdd}
\SetKwFunction{apply}{apply2}
\SetKwFunction{applyone}{apply1}
\SetKwFunction{applysc}{apply2sc}
\SetKwFunction{eval}{eval}
\SetKwFunction{fun}{fun}
\SetKwFunction{leaffun}{leaffun}
\SetKwFunction{funsh}{f}
\SetKwFunction{qtb}{quant\_to\_bool}
\SetKwFunction{leaves}{leaves}
\SetKwFunction{solveone}{solve1}
\SetKwFunction{solvetwo}{solve2}
\SetKwFunction{newvertex}{new\_vertex}
\SetKwFunction{newedge}{new\_edge}
\SetKwFunction{freezevertex}{freeze\_vertex}
\SetKwFunction{setwinner}{set\_winner}
\SetKwFunction{realizability}{realizability}
\SetKwFunction{declarevertex}{declare\_vertex}
\SetKwFunction{refine}{refine}
\SetKwFunction{win}{win\textsubscript{\textsc{o}}}
\SetKwFunction{lose}{win\textsubscript{\textsc{i}}}
\SetKwFunction{dfs}{dfs}
\SetKw{Assert}{assert}
\SetKw{SuchThat}{such that}
\SetKw{Continue}{continue}
\SetKw{Break}{break}
\newcommand{\lIfElse}[3]{\lIf{#1}{#2 \textbf{else}~#3}}

\newcommand{\CA}{\mathcal{A}}

\newcommand{\CE}{\mathcal{E}}
\newcommand{\CF}{\mathcal{F}}
\newcommand{\CG}{\mathcal{G}}
\newcommand{\CI}{\mathcal{I}}
\newcommand{\CO}{\mathcal{O}}
\newcommand{\CS}{\mathcal{S}}
\newcommand{\CV}{\mathcal{V}}
\newcommand{\CP}{\mathcal{P}}
\newcommand{\CQ}{\mathcal{Q}}
\newcommand{\BB}{\mathbb{B}}

\newcommand{\PO}{\textsc{o}\xspace}
\newcommand{\PI}{\textsc{i}\xspace}
\newcommand{\PU}{\textsc{u}\xspace}
\newcommand{\MTBDD}{\mathsf{MTBDD}}
\newcommand{\hashmap}{\mathsf{hashmap}}

\newcommand{\bddlow}{\mathsf{low}}
\newcommand{\bddhigh}{\mathsf{high}}

\newcommand{\lang}{\mathscr{L}}
\newcommand{\term}{\mathsf{term}}
\newcommand{\tr}{\mathsf{tr}}
\newcommand{\subs}{\mathsf{sf}}
\newcommand{\ExtLink}[2]{\href{#1}{\color{blue}#2\,\faExternalLink*}}

\newcommand{\LTLf}{{\texorpdfstring{\ensuremath{\mathsf{LTL_f}}}{LTLf}}\xspace}
\newcommand{\ttrue}{\mathit{tt}}
\newcommand{\ffalse}{\mathit{ff}}
\newcommand{\F}{\mathsf{F}} % eventually
\newcommand{\G}{\mathsf{G}} % always
\newcommand{\X}{\mathsf{X}} % next
\newcommand{\StrongX}{\mathsf{X^{!}}} % strong next
\newcommand{\R}{\mathbin{\mathsf{R}}} % release
\newcommand{\U}{\mathbin{\mathsf{U}}} % until
\newcommand{\limplies}{\rightarrow}
\newcommand{\liff}{\leftrightarrow}
\newcommand{\lxor}{\oplus}

\newcommand{\adl}[2][]{\todo[color=green!40,#1]{ADL: {#2}}}

\definecolor{lime}{HTML}{A6CE39}
\DeclareRobustCommand{\orcidicon}{
	\hspace{-2.5mm}
	\begin{tikzpicture}[baseline={(0,-0.12)}]
	\draw[lime, fill=lime] (0,0)
	circle [radius=0.16]
	node[white] (ID) {{\fontfamily{qag}\selectfont \tiny ID}};
	\draw[white, fill=white] (-0.0625,0.095)
	circle [radius=0.007];
	\end{tikzpicture}
	\hspace{-2.5mm}
      }
\def\orcidID#1{\href{https://orcid.org/#1}{\smash{\orcidicon}}}

\title{Engineering an $\mathsf{LTL_f}$ Synthesis Tool}

\author{Alexandre~Duret-Lutz\inst{1}\orcidID{0000-0002-6623-2512} \and
  Shufang~Zhu\inst{2}\orcidID{0000-0002-5922-8750} \and
  Nir~Piterman\inst{3}\orcidID{0000-0002-8242-5357} \and
  Giuseppe~De~Giacomo\inst{4}\orcidID{0000-0001-9680-7658} \and
  Moshe~Y.~Vardi\inst{5}\orcidID{0000-0002-0661-5773}
}

\authorrunning{A. Duret-Lutz et al.}
\institute{LRE, EPITA, Le Kremlin-Bicêtre, France \and
  University of Liverpool, Liverpool, UK \and
  \!\!\mbox{University\,of\,Gothenburg\,and\,Chalmers\,University\,of\,Technology,\,Gothenburg,\,Sweden}\and
  Sapienza University of Rome, Rome, Italy \and
  Rice University, Houston, Texas, USA}

\begin{document}
\allowdisplaybreaks

\maketitle

\begin{abstract}
  The problem of $\mathsf{LTL_f}$ reactive synthesis is to build a transducer, whose output is based on a history of inputs, such that, for every infinite sequence of inputs, the conjoint evolution of the inputs and outputs has a prefix that satisfies a given $\mathsf{LTL_f}$ specification.

  We describe the implementation of an $\mathsf{LTL_f}$ synthesizer that outperforms existing tools on our benchmark suite.  This is based on a new, direct translation from $\mathsf{LTL_f}$ to a DFA represented as an array of Binary Decision Diagrams (MTBDDs) sharing their nodes. This MTBDD-based representation can be interpreted
  directly as a reachability game that is solved on-the-fly during its construction.
\end{abstract}
\section{Introduction}
\emph{Reactive synthesis} is concerned with synthesizing programs (a.k.a. strategies) for reactive computations (e.g., processes, protocols, controllers, robots) in active environments
\cite{pnueli.89.popl,finkbeiner.16.dsse,ehlers.17.deds}, typically, from temporal logic specifications.
In AI, Reactive Synthesis, which is related to (strong) planning for temporally extended goals in fully observable nondeterministic domains
\cite{cimatti.03.aij,bacchus.98.amai,bacchus.00.aij,calvanese.02.kr,baier.07.icaps,gerevini.09.aij,degiacomo.18.ijcai,camacho.19.icaps},
has been studied with a focus on logics on finite traces such as $\LTLf$~\cite{gabbay.80.popl,baier.06.aaai,degiacomo.13.ijcai,degiacomo.15.ijcai}.
In fact, $\LTLf$ synthesis~\cite{degiacomo.15.ijcai} is one of the two main success stories of reactive synthesis so far (the other being the GR(1) fragment of LTL~\cite{piterman.06.vmcai}), and has brought about impressive advances in scalability~\cite{zhu.17.ijcai,bansal.20.aaai,degiacomo.21.icaps,degiacomo.22.ijcai}.

Reactive synthesis for $\LTLf$ involves the following steps~\cite{degiacomo.15.ijcai}: (1) distinguishing uncontrollable input ($\CI$) and
controllable output ($\CO$) variables in an $\LTLf{}$ specification $\varphi$ of the desired system behavior; (2) constructing a DFA
accepting the behaviors satisfying $\varphi$; (3) interpreting this DFA as a two-player reachability game, and finding a controller winning strategy.
Step (2) has two main bottlenecks: the DFA is worst-case doubly-exponential and its propositional alphabet $\Sigma=2^{\CI\cup\CO}$ is exponential.
The first only happens in the worst case, while the second blow-up -- which we call \emph{alphabet explosion} -- always happens.

Mona~\cite{klarlund.01.tr} addresses the alphabet-explosion problem, which happens also in MSO, by representing a DFA with Multi-Terminal Binary Decision Diagrams (MTBDDs)~\cite{henrikson.95.tacas}.
MTBDDs are a variant of BDDs~\cite{bryant.92.cs} with arbitrary terminal values.  If terminal values encode destination states, an MTBDD can compactly represent all outgoing transitions of a single DFA state. A DFA is represented, through its transition function, as an array of MTBDDs sharing their nodes.

The first $\LTLf$ synthesizer, Syft~\cite{zhu.17.ijcai}, converted \LTLf{} into first-order logic in order to build a MTBDD-encoded DFA with Mona.  Syft then converted this DFA into a BDD representation to solve the reachability game using a symbolic fixpoint computation.  Syft demonstrated that DFA construction is the main bottleneck in $\LTLf$ synthesis, motivating several follow-up efforts.

One approach to effective DFA construction uses compositional
techniques, decomposing the input $\LTLf$ formula into smaller
subformulas whose DFAs can be minimized before being recombined. Lisa~\cite{bansal.20.aaai} decomposes top-level
conjunctions, while Lydia~\cite{giacomo.21.icaps} and LydiaSyft~\cite{favorito.25.tacas}
decompose every operator.

Compositional methods construct the full DFA before synthesis can proceed, limiting their scalability.  On-the-fly approaches~\cite{xiao.21.aaai} construct the DFA incrementally, while simultaneously solving the game, allowing strategies to be found before the complete DFA is built.  The DFA construction may use various techniques.  Cynthia~\cite{degiacomo.22.ijcai} uses Sentential Decision Diagrams (SDDs)~\cite{darwiche.11.ijcai} to
generate all outgoing transitions of a state at once. Alternatively, Nike~\cite{favorito.23.rcra} and MoGuSer~\cite{xiao.2024.vmcai} use a SAT-based method to construct one successor at a time.
% The reachability game is solved by backpropagation during the forward exploration.
The game is solved by forward exploration with suitable backpropagation.
% Space savings (MoGuSer is mentioned later)  Some heuristics can be applied to guide this exploration~\cite{xiao.2024.vmcai}.

\emph{Contributions and Outline} In Section~\ref{sec:translate}, we propose a direct and efficient translation from $\LTLf$ to
MTBDD-encoded DFA (henceforth called MTDFA).  In
Section~\ref{sec:realizability}, we show that given an appropriate ordering of BDD variables, $\LTLf$ realizability can be solved by interpreting the MTBDD nodes of the MTDFA as the vertices of a reachability game, known to be solvable in linear time by backpropagation of the vertices that are winning for the \emph{output} player.
We give a linear-time implementation for solving the game on-the-fly while it is constructed. For more opportunities to abort the on-the-fly construction earlier, we additionally backpropagate vertices that are known to be winning by the \emph{input} player.
We implemented these techniques in two tools
(\ExtLink{https://spot.lre.epita.fr/ltlf2dfa.html}{\texttt{ltlf2dfa}}
and
\ExtLink{https://spot.lre.epita.fr/ltlfsynt.html}{\texttt{ltlfsynt}})
that compare favorably with other existing tools in benchmarks from the $\LTLf$-Synthesis Competition.  To meet space limits, Section~\ref{sec:eval} only reports on the $\LTLf$ realizability benchmark, and we refer readers to our artifact for the other results~\cite{duret.25.zenodo}.\looseness=-1

\vspace*{-1ex}
\section{Preliminaries}
\vspace*{-1ex}
\subsection{Words over Assignments}

A \emph{word over $\sigma$} of length $n$ over an alphabet $\Sigma$ is
a function $\sigma: \{0,1,\ldots,n-1\}\to \Sigma$.  We use $\Sigma^n$
(resp. $\Sigma^\star$ and $\Sigma^+$) to denote the set of words of
length $n$ (resp. any length $n\ge 0$ and $n>0$).  We use $|\sigma|$
to represent the length of a word $\sigma$.
% We use $n=\omega$ for words of infinite size.
For $\sigma\in\Sigma^n$ and $0 \leq i<n$,
% $\sigma(i) \in \Sigma$ denotes the letter at position $i$ of $\sigma$, and
$\sigma(..i)$ denotes the prefix of
$\sigma$ of length $i+1$.
% \sz{Defined $\sigma(i)$.}
% \adl{No objection, but isn't that superfluous since $\sigma$ is defined as a function?}
% \sz{Right, I missed that.}

Let $\CP$ be a finite set of Boolean variables (a.k.a. \emph{atomic
propositions}).  We use $\BB^\CP$ to denote the set of all assignments,
i.e., functions $\CP\to\BB$ mapping variables to values in
$\BB=\{\bot,\top\}$.

Given two disjoint sets of variables $\CP_1$ and $\CP_2$, and two
assignments $w_1\in\BB^{\CP_1}$ and $w_2\in\BB^{\CP_2}$, we use
$w_1\sqcup w_2: (\CP_1\cup \CP_2)\to\BB$ to denote their combination.

In a system modeled using discrete Boolean signals that evolve
synchronously, we assign a variable to each signal, and use a word
$\sigma\in (\BB^\CP)^+$ over assignments of $\CP$ to represent the
conjoint evolution of all signals over time.

We extend $\sqcup$ to such words.  For two words
$\sigma_1\in (\BB^{\CP_1})^n$, $\sigma_2\in (\BB^{\CP_2})^n$ of
length $n$ over assignments that use disjoint sets of variables, we use
$\sigma_1\sqcup\sigma_2 \in (\BB^{\CP_1\cup\CP_2})^n$ to denote a word
such that $(\sigma_1\sqcup\sigma_2)(i) = \sigma_1(i)\sqcup\sigma_2(i)$ for $0\le i<n$.

\subsection{Linear Temporal Logic over Finite, Nonempty Words.}

We use classical $\LTLf$ semantics over nonempty finite
words~\cite{degiacomo.13.ijcai}.

\begin{definition}[$\LTLf$ formulas]
An $\LTLf$ formula
$\varphi$ is built from a set $\CP$ of variables, using the
following grammar where $p\in\CP$, and
$\odot\in\{\land,\lor,\limplies,\liff,...\}$ is any Boolean operator:
$
  \varphi ::= \ttrue\mid\ffalse \mid p \mid \lnot\varphi \mid \varphi\odot\varphi \mid
  \X\varphi \mid \StrongX\varphi \mid \varphi\U\varphi \mid \varphi\R\varphi \mid \G \varphi \mid \F\varphi
  $.\looseness=-1

Symbols $\ttrue$ and $\ffalse$ represent the \emph{true} and \emph{false} $\LTLf$ formulas.
Temporal operators are $\X$ (weak next), $\StrongX$ (strong next), $\U$ (until), $\R$ (release), $\G$ (globally), and $\F$~(finally). $\LTLf(\CP)$ denotes the set of formulas produced by the above grammar. We use $\subs(\varphi)$ to denote the set of subformulas for $\varphi$. A \emph{maximal temporal subformula}\adl{If we drop Def.~\ref{def:propequiv}, we don't need maximal temporal subformulas.} of $\varphi$ is a subformula whose primary operator is temporal and that is not strictly contained within any other temporal subformula of $\varphi$.
%\sz{Moved the definition of $\subs(\varphi)$ here and added the definition of maximal temporal formula, that is later used in the definition of propositional equivalence.}

The satisfaction of a formula $\varphi\in\LTLf(\CP)$ by word $\sigma\in (\BB^\CP)^+$ of length $n>0$ at position $0\le i< n$, denoted $\sigma,i\models\varphi$, is defined as follows.
\vspace*{-.9ex}
\begin{gather*}
  % Group the two aligned blocks in a gather* environment to suppress the space between them.
  \begin{aligned}
    \sigma,i\models \ttrue &\iff i<n &
    \sigma,i\models \X\varphi &\iff (i+1=n)\lor(\sigma,i+1\models \varphi)\\
    \sigma,i\models \ffalse &\iff i=n &
    \sigma,i\models \StrongX\varphi &\iff (i+1<n)\land(\sigma,i+1\models \varphi)\\
    \sigma,i\models p &\iff p\in \sigma(i) &
    \sigma,i\models \F\varphi &\iff \exists j\in [i,n),\,\sigma,j\models \varphi\\
    \sigma,i\models \lnot\varphi &\iff \lnot(\sigma,i\models \varphi) &
    \sigma,i\models \G\varphi &\iff \forall j\in [i,n),\,\sigma,j\models \varphi
  \end{aligned}\\
  \begin{aligned}
  \sigma,i\models \varphi_1\odot\varphi_2 &\iff (\sigma,i\models \varphi_1)\odot(\sigma,i\models \varphi_2)\\
    \sigma,i\models \varphi_1\U\varphi_2 &\iff \exists j{\in} [i,n),(\sigma,j\models \varphi_2)\land(\forall k{\in} [i,j),\,\sigma,k\models \varphi_1)\\
    \sigma,i\models \varphi_1\R\varphi_2 &\iff \forall j{\in} [i,n),(\sigma,j\models \varphi_2)\lor(\exists k{\in} [i,j),\,\sigma,k\models \varphi_1)
  \end{aligned}
\end{gather*}

\vspace*{-.9ex}\noindent
The set of words that satisfy $\varphi\in\LTLf(\CP)$ is $\lang(\varphi)=\{\sigma\in (\BB^\CP)^+ \mid \sigma,0\models \varphi\}$.
\end{definition}

\begin{example}\label{ex:psi}
  Consider the following $\LTLf$ formulas over
  $\CP=\{i_0,i_1,i_2,o_1,o_2\}$:
  $\Psi_1 = \G((i_0\limplies(o_1\liff i_1)) \land ((\lnot i_0)\limplies(o_1\liff i_2)))$, and
  $\Psi_2 = (\G\F o_2)\liff(\F i_0)$.
    If we interpret $i_0,i_1,i_2$ as input signals, and $o_1,o_2$ as output signals,
    formula $\Psi_1$ specifies a 1-bit multiplexer: the value of the signal $o_1$ should
    be equal to the value of either $i_1$ or $i_2$ depending on the setting of $i_0$.
    Formula $\Psi_2$ specifies that the last value of $o_2$ should be $\top$ if and only
    if $i_0$ was $\top$ at some instant.
  \end{example}

%% ADL: We discussed with Guiseppe that this example is too large for the paper, and
%% does not illustrate as many things as the original example.
%
%   \begin{example}\label{ex:psi}
%   Consider the following variables $\CP=\{i_0, i_1,i_2,o_0, o_1,o_2\}$, where we interpret $i_0,i_1,i_2$ as input signals, standing for ``delivered", ``printed on printer 1", ``printed on printed 2" respectively, and = $o_1,o_1,o_2$ as output signals, standing for ``print a job", ``collect from printer 1", and ``collect for printer 2".
%   %%
%   Over $\CP$, consider the following formulas $\LTLf$:
%   $\Psi_0 = \G((o_0 \limplies \lnot o_1\land \lnot o_2) \land (o_1 \limplies \lnot o_0\land \lnot o_1) \land (o_2 \limplies \lnot o_0\land \lnot o_1)))$, expressing that at most one output signal at the time is true;
%   %%
%   $\Psi_1 = \lnot o_1 \land \lnot o_2 \land \lnot o_3$, expressing the values of the output signals at the beginning;
%   %%
%   $\Psi_2 = \G(\X(o_0 \limplies \F(i_1 \lor i_2))) \land \G(i_1 \limplies \X(o_1 \limplies \F(i_0))) \land \G(i_2 \limplies \X(o_2 \limplies \F(i_0)))$, expressing the effects of the outputs on the possible successive inputs;
%   %%
%   $\Psi_3 = \F(i_0)$ (or $\Psi'_3 = \G\F(i_0)$) which states that  we want to see input signal $i_0$ eventually (at last, respectively).
%   %%
%   By using these formulas we give the synthesis specification as
% $\Psi_0\land (\Psi_1\land \Psi_2 \limplies \Psi_3)$.
%
%   \end{example}

\begin{definition}[Propositional Equivalence~\cite{esparza.18.lics}]\label{def:propequiv}
%\np{I think that this is a candidate for space saving. Mention informally propositional equality and move this to appendix?}%
    For $\varphi\in\LTLf(\CP)$, let $\varphi_P$ be the
    Boolean formula obtained from $\varphi$ by replacing every
    maximal temporal subformula $\psi$ by a Boolean variable $x_\psi$.
    Two formulas $\alpha,\beta\in\LTLf(\CP)$ are %said
    \emph{propositionally equivalent}, denoted $\alpha\equiv \beta$, if
    $\alpha_P$ and $\beta_P$ are equivalent Boolean formulas.
  \end{definition}

  \begin{example}
    Formulas $\alpha=(\G b)\lor((\F a)\land(\G b))$ and $\beta=\G b$
    are propositionally equivalent.  Indeed, $\alpha_P=x_{\G b}\lor(x_{\F a}\land x_{\G b})=x_{\G b}=\beta_P$.
  \end{example}

  Note that $\alpha\equiv\beta$ implies $\lang(\alpha)=\lang(\beta)$,
  but the converse is not true in general.  Since $\equiv$ is an
  equivalence relation, we use $[\alpha]_\equiv\in\LTLf(\CP)$ to
  denote some unique representative of the equivalence class of
  $\alpha$ with respect to $\equiv$.

\subsection{\LTLf{} Realizability}

Our goal is to build a tool that decides whether an \LTLf{} formula is \emph{realizable}.

\begin{definition}[\cite{degiacomo.15.ijcai,jacobs.23.tlsf12}]\label{def:realizability}
  Given two disjoint sets of variables $\CI$ (inputs) and $\CO$
  (outputs), a controller is a function $\rho: \CI^*\to\CO$, that
  produces an assignment of output variables given a history of
  assignments of input variables.

  Given a word of $n$ input assignments $\sigma\in(\BB^\CI)^n$, the
  controller can be used to generate a word of $n$ output assignments
  $\sigma_\rho\in(\BB^\CI)^n$.  The definition of $\sigma_\rho$ may
  use two semantics depending on whether we want to the controller to have
  access to the current input assignment to decide the output assignment:
  \begin{description}[nosep]
  \item[Mealy semantics:] $\sigma_\rho(i)=\rho(\sigma(..i))$ for all $0\le i<n$.
  \item[Moore semantics:] $\sigma_\rho(i)=\rho(\sigma(..i-1))$ for all $0\le i<n$.
  \end{description}

  A formula $\varphi\in\LTLf(\CI\cup\CO)$ is said to be \emph{Mealy-realizable} or \emph{Moore-realizable} if there exists a controller $\rho$ such that for any word $\sigma\in(\BB^\CI)^\omega$ there exists a position $k$ such that $(\sigma\sqcup\sigma_\rho)(..k)\in\lang(\varphi)$ using the desired semantics.
\end{definition}

\begin{example}
  Formula $\Psi_1$ (from Example~\ref{ex:psi}) is Mealy-realizable but not Moore-realizable.  Formula $\Psi_2$ is both Mealy and Moore-realizable.
\end{example}

\subsection{Multi-Terminal BDDs}\label{sec:mtbdd}

%\sz{Shall we use $\CQ$ instead of $\CS$? To go smoothly from the terminals of MTBDD to states in DFA.}
%\adl{No, in practice $\CS$ will be a \textbf{pair} in all our uses.  I use $\CS$ on purpose because Appendix~\ref{app:mtbddimpl} explains that the MTBDD implementation knows nothing about our usage to build MTDFAs.  Later we will set $\CS=\CQ\times \BB$ (in MTDFA) or when we start translating LTLf or solving game: $\CS=\LTLf(...)\times \BB$.  In Algorithm~\ref{algo:gamesolving}, $\CQ$ will only contain a subset of the formulas that labels the terminals so $\CS=\LTLf(...)\times\BB$, not $\CQ\times\BB$.}
%\sz{Gotcha!}
Let $\CS$ be a finite set.  Given a finite set of variables
$\CP=\{p_0,p_1,\ldots,p_{n-1}\}$ (that are implicitly ordered by their
index) we use $f:\BB^\CP\to\CS$ to denote a function that maps an
assignment of all those variables to an element of $\CS$.  Given a
variable $p\in\CP$ and a Boolean $b\in\BB$, the function
$f_{p=b}:\BB^{\CP\setminus\{p\}}\to\CS$ represents a generalized
co-factor obtained by replacing $p$ by $b$ in $f$.  %More formally
%$f_{p_i=b}(p_1,p_2,\ldots,p_{i-1},p_{i+1},\ldots,p_{n-1})=f(p_1,p_2,\ldots,p_{i-1},b,p_{i+1},\ldots,p_{n-1})$.
When $\CS=\BB$, a function $f:\BB^\CP\to\BB$ can be encoded into a
Binary Decision Diagram (BDD)~\cite{bryant.86.tc}.  Multi-Terminal
Binary Decision Diagrams
(MTBDDs)~\cite{long.93.bdd,minato.96.vlsi,fujita.97.fmsd,klarlund.01.tr},
also called Algebraic Decision Diagrams
(ADDs)~\cite{bahar.93.iccad,somenzi.15.cudd}, generalize BDDs by
allowing arbitrary values on the leaves of the graph.

A \emph{Multi-Terminal BDD} encodes any function $f:\BB^\CP\to\CS$ as a
rooted, directed acyclic graph.  We use the term \emph{nodes} to refer
to the vertices of this graph.  All nodes in an MTBDD are represented
by triples of the form $(p,\ell,h)$.  In an internal node, $p\in\CP$ and $\ell,h$ point to successors MTBDD nodes called the $\bddlow$ and
$\bddhigh$ links.  The intent is that if $(p,\ell,h)$ is the root of
the MTBDD representing the function $f$, then $\ell$ and $h$ are the
roots of the MTBDDs representing the functions $f_{p=\bot}$ and
$f_{p=\top}$, respectively.  Leaves of the graph, called
\emph{terminals}, hold values in $\CS$.  For consistency with internal
nodes, we represent terminals with a triple of the form
$(\infty,s,\infty)$ where $s\in\CS$.\label{notation:triplet} When
comparing the first elements of different triplets, we assume that
$\infty$ is greater than all variables.  We use $\MTBDD(\CP,\CS)$ to
denote the set of MTBDD nodes that can appear in the representation of
an arbitrary function $\BB^\CP\to\CS$.

Following the classical implementations of BDD
packages~\cite{bryant.86.tc,andersen.99.lecturenotes}, we assume that
MTBDDs are \emph{ordered} (variables of $\CP$ are ordered and visited
in increasing order by all branches of the MTBDD) and \emph{reduced}
(isomorphic subgraphs are merged by representing each triplet only
once, and internal nodes with identical $\bddlow$ and $\bddhigh$ links
are skipped over).  Doing so ensures that each function
$f:\BB^\CP\to\CS$ has a unique MTBDD representation for a given order
of variables.

Given $m\in \MTBDD(\CP,\CS)$ and an assignment $w\in \BB^\CP$, we note
$m(w)$ the element of $\CS$ stored on the terminal of $m$ that
\marginpar{Cf. App.~\ref{app:mtbddeval}} is reached after following
the assignment $w$ in the structure of $m$.  We use $|m|$ to denote
the number of MTBDD nodes that can be reached from $m$.

Let $m_1\in\MTBDD(\CP,\CS_1)$ and $m_2\in\MTBDD(\CP,\CS_2)$ be two
MTBDD nodes representing functions $f_i:\BB^\CP\to\CS_i$, and let
$\odot:\CS_1\times\CS_2\to\CS_3$, be a binary operation.  One can
easily construct $m_3\in\MTBDD(\CP,\CS_3)$ representing the function
$f_3(p_0,\ldots,p_{n-1})=f_1(p_0,\ldots,p_{n-1})\odot
f_2(p_0,\ldots,p_{n-1})$, by generalizing the \apply function
typically found in BDD
libraries~\cite{fujita.97.fmsd}.\marginpar{Cf. App.~\ref{app:mtbddapply}}
We use $m_1\odot m_2$ to denote the MTBDD that results from this
construction.

For $m\in \MTBDD(\CP,\CS)$ we use $\leaves(m)\subseteq\CS$ to denote
the elements of $\CS$ that label terminals reachable from $m$.  This
set can be computed in $\Theta(|m|)$.\marginpar{Cf. App.~\ref{app:leaves}.}

\subsection{MTBDD-Based Deterministic Finite Automata}

We now define an MTBDD-based representation of a DFA with a propositional
alphabet, inspired by Mona's DFA representation~\cite{henrikson.95.tacas,klarlund.01.tr}.

\vspace*{-.5ex}
\begin{definition}[MTDFA]
  An MTDFA is a tuple $\CA=\langle \CQ, \CP, \iota, \Delta\rangle$, where $\CQ$ is a finite set of \emph{states}, $\CP$ is a finite (and ordered) set of
  \emph{variables}, $\iota\in \CQ$ is the initial state,
  $\Delta: \CQ\to \MTBDD(\CP,\CQ\times\BB)$ represents the set of
  outgoing transitions of each state.
For a word $\sigma\in(\BB^\CP)^\star$ of length $n$, let
$(q_i,b_i)_{0\le i \le n}$ be a sequence of pairs defined recursively
as follows: $(q_0,b_0)=(\iota,\bot)$, and for $0<i\le|\sigma|$,
$(q_i,b_i)=\Delta(q_{i-1})(\sigma(i-1))$ is the pair reached by
evaluating assignment $\sigma(i-1)$ on $\Delta(q_{i-1})$.
The word $\sigma$ is accepted by $\CA$ iff $b_n=\top$.
The language of $\CA$, denoted $\lang(A)$, is the set of words
accepted by $\CA$.
\end{definition}
\vspace*{-.8ex}

\begin{figure}[t]
  \begin{tikzpicture}[scale=.74]
    \node[root,left] at (0,0) (r0) {$\Psi_1 \land ((\G\F o_2){\liff}(\F i_0))$};
    \node[left] at (r0.west) (init) {$\iota=$};
    \node[root,left] at (0,-1) (r1) {$\Psi_1 \land ((\F o_2\land \G\F o_2){\liff}(\F i_0))$};
    \node[root,left] at (0,-2) (r2) {$\ffalse$};
    \node[root,left] at (0,-3) (r3) {$\Psi_1 \land (\G\F o_2)$};
    \node[root,left] at (0,-4) (r4) {$\Psi_1 \land ((\G\F o_2)\land(\F o_0))$};
    \node[inode] at (1, -0.5) (i00) {$i_0$};
    \node[inode] at (1, -3.5) (i01) {$i_0$};
    \node[inode] at (2.2, -0.5) (i20) {$i_2$};
    \node[inode] at (2.2, -3) (i10) {$i_1$};
    \node[inode] at (2.2, -4) (i21) {$i_2$};
    \node[inode] at (3.4, -0) (o10) {$o_1$};
    \node[inode] at (3.4, -1) (o11) {$o_1$};
    \node[inode] at (3.4, -3) (o12) {$o_1$};
    \node[inode] at (3.4, -4) (o13) {$o_1$};
    \node[inode] at (4.6, -0) (o20) {$o_2$};
    \node[inode] at (4.6, -4) (o21) {$o_2$};
    \node[termn,right] at (5.6,0) (t0) {$\Psi_1 \land ((\G\F o_2){\liff}(\F i_0))$};
    \node[terma,right] at (5.6,-1) (t1) {$\Psi_1 \land ((\F o_2\land \G\F o_2){\liff}(\F i_0))$};
    \node[termn,right] at (5.6,-2) (t2) {$\ffalse$};
    \node[terma,right] at (5.6,-3) (t3) {$\Psi_1 \land (\G\F o_2)$};
    \node[termn,right] at (5.6,-4) (t4) {$\Psi_1 \land ((\G\F o_2)\land(\F o_2))$};
    \draw[high] (i00) -> (i10);
    \draw[low]  (i00) -> (i20);
    \draw[high] (i01) -> (i10);
    \draw[low]  (i01) -> (i21);
    \draw[high] (i20) -> (o10);
    \draw[low]  (i20) -> (o11);
    \draw[high] (i10) -> (o13);
    \draw[low]  (i10) -> (o12);
    \draw[high] (i21) -> (o13);
    \draw[low]  (i21) -> (o12);
    \draw[high] (o10) -> (o20);
    \draw[low]  (o10) -> (t2);
    \draw[high] (o11) -> (t2);
    \draw[low]  (o11) -> (o20);
    \draw[high] (o12) -> (t2);
    \draw[low]  (o12) -> (o21);
    \draw[high] (o13) -> (o21);
    \draw[low]  (o13) -> (t2);
    \draw[high] (o20) -> (t0);
    \draw[low]  (o20) -> (t1.west);
    \draw[high] (o21) -> (t3.west);
    \draw[low]  (o21) -> (t4);
    \draw[rlink] (r0.east) -> (i00);
    \draw[rlink] (r1.east) -> (i00);
    \draw[rlink] (r2.east) -> (t2);
    \draw[rlink] (r3.east) -> (i01);
    \draw[rlink] (r4.east) -> (i01);
  \end{tikzpicture}\vspace*{-3mm}
  \caption[An MTDFA]{An MTDFA where $\CP=\{i_0,i_1,i_2,o_0,o_1\}$ and $\CQ\subseteq\LTLf(\CP)$.  Following classical BDD representations a BDD node $(p,\ell,h)$ is represented by
    \begin{tikzpicture}[baseline=(i.base)]
      \node[inode] at (0,0) (i) {$p$};
      \node[overlay] at (0.75,0.15) (l) {$\ell$};
      \node[overlay] at (0.75,-0.15) (h) {$h$};
      \draw[low,overlay] (i) -> (l);
      \draw[high] (i) -> (h);
    \end{tikzpicture}~~. A terminal $(\infty,(\alpha,b),\infty)$ is represented by
    \tikz[baseline=(t.base)]\node[termn](t){$\alpha$}; if $b=\bot$, or
    \tikz[baseline=(t.base)]\node[terma](t){$\alpha$}; if $b=\top$.
    Finally, MTBDD $m=\Delta(\alpha)$ representing the successors of state $\alpha$
    is indicated with
    \begin{tikzpicture}[baseline=(r.base)]
      \node[root] at (0,0) (r) {$\alpha$};
      \node at (1,0) (b) {$m$};
      \draw[rlink] (r) -> (b);
    \end{tikzpicture}\!\!.  Subformula $\Psi_1$ abbreviates $\G((i_0\limplies(o_1\liff i_1)) \land ((\lnot i_0)\limplies(o_1\liff i_2)))$.\label{fig:mtdfa}}
    \vspace*{-4mm}
\end{figure}

\begin{example}
  Figure~\ref{fig:mtdfa} shows an MTDFA where
  $\CQ\subseteq\LTLf(\{i_0,i_1,i_2,o_1,o_2\})$.  The set of states
  $\CQ$ are the dashed rectangles on the left.  For each such a state
  $q\in\CQ$, the dashed arrow points to the MTBDD node representing
  $\Delta(q)$.  The MTBDD nodes are shared between all states.  If,
  starting from the initial state $\iota$ at the top-left, we
  read the assignment
  $w=(i_0{\to}\top,i_1{\to}\top,i_2{\to}\top,o_1{\to}\top,o_2{\to}\top)$, we
  should follow only the $\bddhigh$ links (plain arrows) and we reach
  the $\Psi_1\land(\G\F o_2)$ accepting terminal.  If we read
  this assignment a second-time, starting this time from state
  $\Psi_1\land(\G\F o_2)$ on the left, we reach the same accepting
  terminal.  Therefore, non-empty words of the form $www\ldots w$ are
  accepted by this automaton.
\end{example}

An MTDFA can be regarded as a semi-symbolic representation of a DFA
over propositional alphabet.\marginpar{Cf.~App.~\ref{app:simplified}}
From a state $q$ and reading the assignment $w$, the automaton jumps
to the state $q'$ that is the result of computing
$(q',b)=\Delta(q)(w)$.  The value of $b$ indicates whether that
assignment is allowed to be the last one of the word being read.  By
definition, an MTDFA cannot accept the empty word.

MTDFAs are compact representations of DFAs, because the MTBDD
representation of the successors of each state can share their common
nodes.  Boolean operations can be implemented over MTDFAs, with the
expected semantics, i.e.,
$\lang(\CA_1\odot\CA_2) = \{\sigma\in (\BB^\CP)^+ \mid (\sigma \in
\lang(\CA_1)) \odot (\sigma \in \lang(\CA_2))\}$.\marginpar{Cf. App.~\ref{app:mtdfaops}}

\section{Translating $\LTLf$ to MTBDD and MTDFA}\label{sec:translate}

%% ADL: The commented sentence bellow contradicts the introduction
%% which does not mention "alternating automata" as an approach for
%% obtaining a DFA.  Is there any tool that implement that ``standard''
%% method?
%%
%The standard translation from \LTLf to DFA goes through
%alternating automata to nondeterministic automata to deterministic
%automata~\cite{degiacomo.13.ijcai}.

This section shows how to directly transform a formula
$\varphi\in\LTLf(\CP)$ into an MTDFA
$\CA_\varphi=\langle \CQ, \CP, \varphi, \Delta\rangle$ such that
$\lang(\varphi)=\lang(\CA_\varphi)$.  The translation is reminiscent
of other translations of \LTLf to
DFA~\cite{degiacomo.13.ijcai,degiacomo.22.ijcai}, but it
leverages the fact that MTBBDs can provide a normal form for \LTLf
formulas.

The construction maps states to $\LTLf$ formulas, i.e.,
$\CQ\subseteq\LTLf(\CP)$.  Terminals appearing in the MTBDDs of
$\CA_\varphi$ will be labeled by pairs
$(\alpha,b)\in\LTLf(\CP)\times \BB$, so we use
$\term(\alpha,b)=(\infty,(\alpha,b),\infty)$ to shorten the notation
from Section~\ref{notation:triplet}.

The conversion from $\varphi$ to $\CA_\varphi$ is based on the
function $\tr: \LTLf(\CP)\to \MTBDD(\CP,\LTLf(\CP)\times\BB)$ defined
inductively as follows:\\[-1.8em]
\begin{align*}
  \tr(\ffalse)&=\term(\ffalse,\bot)&
  \tr(\X\alpha)&=\term(\alpha,\top)\\
  \tr(\ttrue)&=\term(\ttrue,\top)&
  \tr(\StrongX\alpha)&=\term(\alpha,\bot)\\
  \tr(p)&=(p,\term(\ffalse,\!\bot),\term(\ttrue,\!\top))\text{~for~}p{\,\in\,}\CP &
\tr(\lnot \alpha)&=\lnot\tr(\alpha)\\
\tr(\alpha\odot\beta)&=\tr(\alpha)\odot\tr(\beta)\text{~for any~}\odot\in\mathrlap{\{\land,\lor,\limplies,\liff,\lxor\}}\\
  \tr(\alpha\U\beta)&=\tr(\beta)\lor(\tr(\alpha)\land\term(\alpha\U\beta,\bot)) &
  \tr(\F \alpha)&=\tr(\alpha)\lor\term(\F\alpha,\bot)\\
\tr(\alpha\R\beta)&=\tr(\beta)\land(\tr(\alpha)\lor\term(\alpha\R\beta,\top))&
\tr(\G \alpha)&=\tr(\alpha)\land\term(\G\alpha,\top)
\end{align*}\\[-1.8em]
Boolean operators that appear to the right of the equal sign are
applied on MTBDDs as discussed in Section~\ref{sec:mtbdd}. % and
% Algorithm~\ref{algo:apply2}.
Terminals in $\LTLf(\CP)\times\BB$ are combined with:
$(\alpha_1,b_1)\odot(\alpha_2,b_2) =
([\alpha_1\odot\alpha_2]_\equiv,b_1\odot b_2)$ and $\lnot(\alpha,b)=([\lnot\alpha]_\equiv,\lnot b)$.

\pagebreak[2]

\begin{theorem}\label{th:translation}
  For $\varphi\in\LTLf(\CP)$, let
  $\CA_\varphi=\langle \CQ, \CP, \iota, \Delta\rangle$ be the MTDFA
  obtained by setting $\iota=[\varphi]_\equiv$, $\Delta=\tr$, and
  letting $\CQ$ be the smallest subset of $\LTLf(\CP)$ such that
  $\iota\in \CQ$, and such that for any $q\in\CQ$ and for any
  $(\alpha,b)\in\leaves(\Delta(q))$, then $\alpha\in\CQ$.  With this
  construction, $|\CQ|$ is finite and
  $\lang(\varphi)=\lang(\CA_\varphi)$.
\end{theorem}
%\vspace*{-1.5ex}
\begin{proof} (sketch) By definition of $\tr$, $\CQ$ contains
  only Boolean combinations of subformulas of $\varphi$. Propositional
  equivalence implies that the number of such combinations is
  finite: $|\CQ|\le 2^{2^{|\subs(\varphi)|}}$.  The language equivalence
  follows from the definition of $\LTLf$, and
  from some classical $\LTLf$ equivalences.  For instance
  the rule for $\tr(\alpha\U\beta)$ is based on the equivalence
  $\lang(\alpha\U\beta)=\lang(\beta\lor(\alpha\land\StrongX(\alpha\U\beta)))$.
\end{proof}
%\vspace*{-1.5ex}
\begin{example}
  Figure~\ref{fig:mtdfa} is the MTDFA for formula $\Psi_1\land\Psi_2$,
  presented in Example~\ref{ex:psi}.  Many more examples can be found
  \marginpar{Also App.~\ref{app:tryonline}}
  in the associated artifact~\cite{duret.25.zenodo}.
\end{example}

The definition of $\tr(\cdot)$ as an MTBDD representation of the set
of successors of a state can be thought as a symbolic representation
of Antimirov's linear forms~\cite{antimirov.96.tcs} for DFA with
propositional alphabets.
%\adl{Some people at CIAA love
%  derivatives and partial derivatives, so I think the analogy should
%  help.}
Antimirov presented linear forms as an efficient way to construct all
(partial) derivatives at once, without having to iterate over the
alphabet.  For \LTLf, \emph{formula
  progressions}~\cite{degiacomo.22.ijcai} are the equivalent of
Brozozowski derivatives~\cite{brzozowski.64.acm}.  Here,
$\tr(\cdot)$ computes all formulas progressions at once, without
having to iterate over an exponential number of assignments.\looseness=-1

Finally, note that while this construction works with any order for
$\CP$, different orders might produce a different number of MTBDD
nodes.

\paragraph{Optimizations}\label{sec:trans-optims}

The previous definitions can be improved in several ways.

Our implementation of MTBDD actually supports terminals that are the
Boolean terminals of standard BDDs as well as the terminals used so
far.  So we are actually using
$\MTBDD(\CP,(\LTLf(\CP)\times \BB)\cup\BB)$, and we encode
$\term(\ffalse,\bot)$ and $\term(\ttrue,\top)$ directly as $\bot$ and
$\top$ respectively.  With those changes, \apply may be modified to
shortcut the recursion depending on the values of $m_1$, $m_2$, and
$\odot$.
% \np{Shouldn't this be $\land$?} ADL: Indeed!
For instance if $\odot=\land$ and $m_1=\top$, then $m_2$ can
be returned immediately.  \marginpar{Cf. App~\ref{app:apply2sc}} Such
shortcuts may be implemented for $\MTBDD(\CP,\CS\cup\BB)$ regardless
of the nature of $\CS$, so our implementation of MTBDD operations is
independent of $\LTLf$.

When combining terminals during the computation of $\tr$, one has to
compute the representative formula $[\alpha_1\odot\alpha_2]_\equiv$.
This can be done by converting ${\alpha_1}_P$ and ${\alpha_2}_P$ into
BDDs, keeping track of such conversions in a hash table.  Two
propositionally equivalent formulas will have the same BDD
representation.  While we are looking for a representative formula, we
can also use the opportunity to simplify the formula at hand.  We
use the following very simple rewritings, for patterns that occur
naturally in the output of $\tr$:\\
$(\alpha\U\beta)\lor \beta\leadsto\alpha\U\beta$, \hfill
$(\alpha\R\beta)\land \beta\leadsto\alpha\R\beta$, \hfill
$(\F\beta)\lor \beta\leadsto\F\beta$, \hfill
$(\G\beta)\land \beta\leadsto\G\beta$.

Once $\CA_\varphi$ has been built, %, we can look at the different
% values of $\Delta$ to detect states with identical successors.
two
states $q,q'\in\CQ$ such that $\Delta(q)=\Delta(q')$ can be merged by
replacing all occurrences of $q'$ by $q$ in the leaves of $\Delta$.
% $\Delta(s)$ for any $s\in\CQ$.

\begin{example}\label{ex:merge}
  The automaton from Figure~\ref{fig:mtdfa}
  \marginpar{Cf.~App.~\ref{app:simplified}~\&~\ref{app:tryonline}}
  has two pairs of states
  that can be merged.  However, if the rule
  $(\G\beta)\land \beta\leadsto\G\beta$ is applied during the
  construction, then the occurrence of $(\G\F o_2)\land (\F o_2)$ will
  already be replaced by $\G\F o_2$, producing the simplified
  automaton without requiring any merging.
\end{example}

\section{Deciding \LTLf{} Realizability}\label{sec:realizability}

\LTLf{} realizability (Def.~\ref{def:realizability}) is solved by
reducing the problem to a two-player reachability game where one
player decides the input assignments and the other player decides the
output
assignments~\cite{degiacomo.15.ijcai}. Section~\ref{sec:reachgame}
presents reachability games and how to interpret the MTDFA as a
reachability game, and Section~\ref{sec:realizability-otf} shows how
we can solve the game on-the-fly while constructing it.

\subsection{Reachability Games \& Backpropagation}\label{sec:reachgame}

\begin{definition}[Rechability Game]
  A Reachability Game is
  $\CG=\langle \CV=\CV_\PO \biguplus \CV_\PI,\linebreak[3] \CE,
  \CF_\PO\rangle$, where $\CV$
  is a finite set of \emph{vertices} partitioned to
  player \emph{output} (abbreviated \PO) and player \emph{input} (abbreviated \PI), $\CE\subseteq \CV\times \CV$ is a finite set of \emph{edges}, and $\CF_\PO\subseteq \CV$ is the set of target states.  Let $\CE(v)=\{ (v,v') ~|~ (v,v')\in \CE\}$. This graph is also referred to as the game \emph{arena}.
  %We have
  %$\CF_\PO\subseteq \CV_\PI$ and $\CE(v)=\emptyset$ for every
  %$v\in \CF_\PO$.\adl{I don't have these restrictions about $\CF_\PO$ in the code.  Is it needed?}

%A strategy for player \PO is $\sigma :\CV \rightarrow \mathbb{N}\cup \{\infty\}$ such that for every $v\in \CV$ such that $\sigma(v)\neq\infty$ we have (a) if $v\in \CV\setminus \CF$ then there is some $v'\in \CE(v)$ such that $\sigma(v')<\sigma(v)$ and (b) if $v\in \CV_\PI$ then for every $v'\in \CE(v)$ we have $\sigma(v')<\sigma(v)$.
%A vertex $v$ is winning for \PO if there exists a strategy $\sigma$ such that  $\sigma(v)<\infty$ .
%Notice that every dead-end that is not in $\CF$ is not winning.
A strategy for player \PO is  a cycle-free subgraph $\langle W,\sigma\rangle
\subseteq \langle \CV,\CE\rangle$ such that (a) for every $v\in W$ we have $v\in \CF_\PO$ or $\CE(v)\cap \sigma \neq \emptyset$ and (b) if $v\in W\cap \CV_\PI$ then $\CE(v)\subseteq \sigma$.
A vertex $v$ is winning for \PO if $v\in W$ for some strategy $\langle W,\sigma \rangle$.
\end{definition}
Such a reachability game can be solved by backpropagation identifying
the maximal set $W$ in a strategy.  Namely, start from $W=\CF_{\PO}$.
Then $W$ is iteratively augmented with every vertex in $\CV_{\PO}$
that has some edge to $W$, and every (non dead-end) vertex in
$\CV_{\PI}$ whose edges all lead to $W$.  At the end of this
backpropagation, which can be performed in linear time~\cite[Theorem 3.1.2]{gradel.07.book}, every vertex in $W$ is winning for $\PO$, and every vertex outside $W$ is losing for $\PO$.  Notice that every dead-end that
is not in $\CF_\PO$ cannot be winning.  It follows that we can
identify some (but not necessarily all) vertices that are losing by setting $L$ as the set of all dead-ends and adding to $L$ every $\CV_{\PI}$ vertex that has some edge to $L$ and every $\CV_{\PO}$ vertex whose edges all lead to $L$.

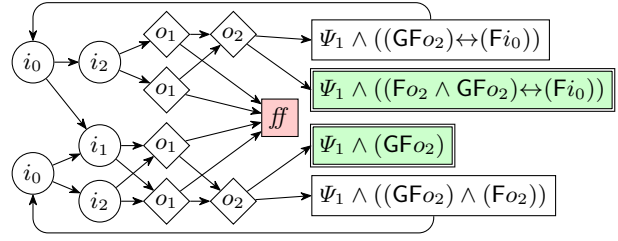
\begin{figure}[t]
  \centering
  \begin{tikzpicture}[scale=.74]
    \node[inode] at (1, -1) (i00) {$i_0$};
    \node[inode] at (1, -3) (i01) {$i_0$};
    \node[inode] at (2.2, -1) (i20) {$i_2$};
    \node[inode] at (2.2, -2.5) (i10) {$i_1$};
    \node[inode] at (2.2, -3.5) (i21) {$i_2$};
    \node[control] at (3.4, -0.5) (o10) {$o_1$};
    \node[control] at (3.4, -1.5) (o11) {$o_1$};
    \node[control] at (3.4, -2.5) (o12) {$o_1$};
    \node[control] at (3.4, -3.5) (o13) {$o_1$};
    \node[control] at (4.6, -0.5) (o20) {$o_2$};
    \node[control] at (4.6, -3.5) (o21) {$o_2$};
    \node[termn,right] at (6,-0.6) (t0) {$\Psi_1 \land ((\G\F o_2){\liff}(\F i_0))$};
    \node[win,terma,right] at (6,-1.5) (t1) {$\Psi_1 \land ((\F o_2\land \G\F o_2){\liff}(\F i_0))$};
    \node[lose,termn,right] at (5.1,-2) (t2) {$\ffalse$};
    \node[win,terma,right] at (6,-2.5) (t3) {$\Psi_1 \land (\G\F o_2)$};
    \node[termn,right] at (6,-3.4) (t4) {$\Psi_1 \land ((\G\F o_2)\land(\F o_2))$};
    \draw[->] (i00) -- (i10);
    \draw[->] (i00) -- (i20);
    \draw[->] (i01) -- (i10);
    \draw[->] (i01) -- (i21);
    \draw[->] (i20) -- (o10);
    \draw[->] (i20) -- (o11);
    \draw[->] (i10) -- (o13);
    \draw[->] (i10) -- (o12);
    \draw[->] (i21) -- (o13);
    \draw[->] (i21) -- (o12);
    \draw[->] (o10) -- (o20);
    \draw[->] (o10) -- (t2);
    \draw[->] (o11) -- (t2);
    \draw[->] (o11) -- (o20);
    \draw[->] (o12) -- (t2);
    \draw[->] (o12) -- (o21);
    \draw[->] (o13) -- (o21);
    \draw[->] (o13) -- (t2);
    \draw[->] (o20) -- (t0.west);
    \draw[->,rounded corners=2mm] (t0.north) -- ++(0,.3) -| (i00);
    \draw[->] (o20) -- (t1.west);
    \draw[->] (o21) -- (t3.west);
    \draw[->] (o21) -- (t4.west);
    \draw[->,rounded corners=2mm] (t4.south) -- ++(0,-.3) -| (i01);
  \end{tikzpicture}
  \caption[A game]{Interpretation of the MTDFA of Figure~\ref{fig:mtdfa} as a game
    with $\CI=\{i_0,i_1,i_2\}$, $\CO=\{o_1,o_2\}$.  Each MTBDD \emph{node} of the MTDFA is
    viewed as a \emph{vertex} of the game, with terminal of the form $(\alpha,\bot)$
    looping back to $\Delta(\alpha)$.  Player \PO decides where to go
    from diamond and rectangular vertices and wants to reach the \colorbox{green!20}{green vertices} corresponding to accepting terminals.
    Player \PI decides where to go from round vertices and wants to reach
    \colorbox{red!20}{$\ffalse$} or avoid green vertices.  %Rectangular vertices
    %representing terminals are played by \PO.%
    % , but it does not really matter as they allow a unique (forced) move when
    % they are not winning for one player.
    \label{fig:game}}
    \vspace*{-7mm}
\end{figure}

Let $\CA_\varphi=\langle\CQ,\CI\uplus\CO,\iota,\Delta\rangle$ be a translation of $\varphi\in\LTLf(\CI\uplus\CO)$ (per Th.~\ref{th:translation}) such that variables of $\CI$ appear before $\CO$ in the MTBDD encoding of $\Delta$.

\begin{definition}[Realizability Game]\label{def:realgame}
We define the reachability game $\CG_\varphi=\langle\CV=\CV_\PI\uplus\CV_\PO,\CE,\CF_\PO\rangle$ in which $\CV\subseteq\MTBDD(\CI\uplus\CO)$ corresponds the set of nodes that appear in the MTBDD encoding of $\Delta$. $\CV_\PO$ contains all nodes $(p,\ell,h)$ such that $p\in\CO$ or $p=\infty$ (terminals), and $\CV_\PI$ contains those with $p\in\CI$.  The edges $\CE$ follows the structure of $\Delta$, i.e., if $\CA_\varphi$ has a node $r=(p,\ell,h)$, then $\{(r,\ell),(r,h)\}\subseteq \CE$.   Additionally, for any terminal $t=(\infty,(\alpha,\bot),\infty)$ such that $\alpha\ne\ffalse$, $\CE$ contains the edge $(t,\Delta(\alpha))$.  Finally, $\CF_\PO$ is the set of accepting terminals, i.e., nodes of the form $(\infty,(\alpha,\top),\infty)$.
\end{definition}
\vspace*{-3ex}
\begin{theorem}\label{th:encode}
  Vertex $\Delta(\iota)$ is winning for \PO in $\CG_\varphi$ iff
  $\varphi$ is Mealy-realizable.
\end{theorem}
\vspace*{-1ex}
Moore realizability can be checked similarly by changing the order of
$\CI$ and $\CO$ in the MTBDD encoding of $\Delta$.
\vspace*{-1ex}
\begin{example}
  Figure~\ref{fig:game} shows how to interpret the MTDFA of
  Figure~\ref{fig:mtdfa} as a game, by turning each MTBDD node into a
  game vertex.  The player owning each vertex is chosen according to
  the variable that labels it.  Vertices corresponding to accepting
  terminals become winning targets for the output player, so the game
  stops once they are reached.  Solving this game will find every
  internal node as winning for $\PO$, so the corresponding formula is
  Mealy-realizable.
\end{example}

The difference with DFA games~\cite{degiacomo.15.ijcai,degiacomo.22.ijcai,favorito.23.rcra,xiao.2024.vmcai} is that instead of having player \PI select all input signals at once, and then player \PO select all output signals at once, our game proceeds by selecting one signal at a time. Sharing nodes that represent identical partial assignments contributes to the scalability of our approach.\looseness=-1

\subsection{Solving Realizability On-the-fly}\label{sec:realizability-otf}

We now show how to construct and solve $\CG_\varphi$ on-the-fly, for better efficiency.  The construction is easier to study in two parts: (1) the on-the-fly solving of reachability games, based on backpropagation, and (2) the incremental construction of $\CG_\varphi$, done with a forward exploration of a subset of the MTDFA for $\varphi$.\looseness=-1

%Algorithm~\ref{algo:gamesolving} provides function for creating a reachability game by adding vertices and edges, and solving it.

%, and how to solve such a game during its construction %(Algorithm~\ref{algo:gamesolving}).  We now combine everything to provide an on-the-fly algorithm for solving realizability. To do so, we do not construct the MTDFA.  Instead, Algorithm~\ref{algo:realizability} builds the game from the \LTLf{} formula incrementally, by essentially translating
%one state $\alpha$ of the MTDFA at a time, and creating game vertices for each MTBDD nodes of $\tr(\alpha)$ using the functions from Algorithm~\ref{algo:gamesolving}.

%we
%The construction in Section~\ref{sec:reachgame} improves on the approach of %using Mona for this construction \cite{zhu.17.ijcai} by constructing the %MTDFA directly, rather than through translation to FOL.
%Here we show that we can combine the direct construction of the MTDFA with on-%the-fly solution of the realizability game.
%That is, we construct the states of the MTDFA incrementally as we evaluate the reachability game for the initial state.

Algorithm~\ref{algo:gamesolving} presents the first part: a set of functions for constructing a game arena incrementally, while performing the
\marginpar{Cf. App.~\ref{app:tryonline}}
linear-time backpropagation algorithm on-the-fly.  At all points
during this construction, the winning status of a vertex
($\mathit{winner}[x]$) will be one of $\PO$ (player \PO can force the
play to reach $\CF_\PO$, i.e., the vertex belongs to $W$), $\PI$
(player \PI can force the play to avoid $\CF_\PO$, i.e., the vertex
belongs to $L$), or $\PU$ (undetermined yet), and the algorithm will
backpropagate both $\PO$ and $\PI$.  At the end of the construction,
all vertices with status $\PU$ will be considered as winning for \PI.
Like in the standard algorithm for solving reachability
games~\cite[Th.~3.1.2]{gradel.07.book} each state uses a counter
($\mathit{count}$, lines~\ref{ln:cntwrite},\ref{ln:cntupdate}) to
track the number of its undeterminated successors.  When a vertex $x$
is marked as winning for player $w$ by calling $\setwinner{x,w}$, an
undeterminated predecessor $p$ has its counter decreased
(line~\ref{ln:cntupdate}), and $p$ can be marked as winning for $w$
(line~\ref{ln:propagate}) if either vertex $p$ is owned by $w$ (player
$w$ can choose to go to $x$) or the counter dropped to $0$ (meaning
that all choices at $p$ were winning for $w$).%\looseness=-1

To solve the game while it is constructed, we \emph{freeze} vertices.
A vertex should be frozen after
all its successors have been introduced with \newedge.  The counter
dropping to $0$ is only checked on frozen vertices
(lines~\ref{ln:cntchk1},~\ref{ln:cntchk2}) since it is only meaningful
if all successors of a vertex are known.\looseness=-1

  \LinesNumbered
\begin{algorithm}[tbp]
  \caption[Solving  reachability game on-the-fly]{API for
    solving a reachability game on-the-fly.  Construct the game
    arena with \newvertex and \newedge.  Once all successors of a
    vertex have been connected, call \freezevertex.
    Call \setwinner at any point to designate vertices winning for
    one player.  %The winning player of each
    %vertex will be updated in $\mathit{winner}$ while the game arena is
    %constructed.
    \vspace*{-5mm}
\label{algo:gamesolving}}
  \Var{$\mathit{owner}[]$\tcp*[r]{map each vertex to one of $\{\PO,\PI\}$}}
  \Var{$\mathit{pred}[]$\tcp*[r]{map vertices to sets of predecessor vertices}}
  \Var{$\mathit{count}[]$\tcp*[r]{map vertices to \# of undeterminated successors}}
  \Var{$\mathit{winner}[]$\tcp*[r]{map vertices to one of $\{\PO,\PI,\PU\}$}}
  \Var{$\mathit{frozen}[]$\tcp*[r]{map vertices to their frozen status (a Boolean)}}
  \Fn(\tcp*[h]{new vertex owned by $\mathit{own}$}){\newvertex{$x\in\CV,\mathit{own}\in\{\PO,\PI\}$}}{
    $\mathit{owner}[x]\gets \mathit{own}$;
    $\mathit{pred}[x]\gets \emptyset$;
    $\mathit{count}[x]\gets 0$;
    $\mathit{frozen}[x]\gets \bot$\;
    $\mathit{winner}[x]\gets \PU$\tcp*[l]{undeterminated winner}
  }
  \Fn{\newedge{$\mathit{src}\in\CV,\mathit{dst}\in\CV$}}{
    \Assert($\mathit{frozen}[\mathit{src}]=\bot$)\;
    \uIf{$\mathit{winner}[\mathit{dst}]=\PU$}{
      $\mathit{count}[\mathit{src}]\gets\mathit{count}[\mathit{src}]+1$;\label{ln:cntwrite}
      $\mathit{pred}[\mathit{dst}]\gets \mathit{pred}[\mathit{dst}]\cup \{\mathit{src}\}$\;
    }\lElseIf{$\mathit{winner}[\mathit{dst}] = \mathit{owner}[\mathit{src}]$}{
      \setwinner{$\mathit{src}, \mathit{owner}[\mathit{src}]$}%
    }
    \tcp{ignore the edge otherwise, it will never be used}
  }
  \Fn(\tcp*[h]{promise not to add more successors}){\freezevertex{$\mathit{x}\in\CV$}}{
    $\mathit{frozen}[x]\gets \top$%
    \tcp*[r]{next line, we assume $\lnot\PI=\PO$ and $\lnot\PO=\PI$}
    \lIf{$\mathit{winner}[x]=\PU\land \mathit{count}[n]=0$\label{ln:cntchk1}}{
      \setwinner{$x, \lnot \mathit{owner}[x]$}
    }
  }
  \Fn(\tcp*[h]{with linear backprop.}){\setwinner{$\mathit{x}\in\CV,w\in\{\PO,\PI\}$}}{
    %% Iterative version
    % \Assert($\mathit{winner}[x] = \PU$);
    % $\mathit{winner}[x] \gets \mathit{win}$;
    % $\mathit{todo}\gets\{x\}$\;
    % \While{$\mathit{todo}\ne\emptyset$}{
    %   $y \gets \mathit{todo}.\mathsf{pop\_any}()$;
    %   $w \gets \mathit{winner}[y]$\;
    %   \ForEach{$p\in\mathit{pred}[y]$ \SuchThat $\mathit{winner}[p]=\PU$}{
    %    $\mathit{count}[p] \gets \mathit{count}[p] - 1$\;
    %    \If{$\mathit{owner}[p] = w \lor (\mathit{count}[p] = 0\land \mathit{frozen}[p])$}{
    %      $\mathit{winner}[p]\gets w$;
    %      $\mathit{todo}\gets \mathit{todo}\cup \{p\}$\;
    %    }
    %    }
    %% The recursive version takes fewer lines
    \Assert($\mathit{winner}[x] = \PU$);
    $\mathit{winner}[x] \gets w$\;
    \ForEach{$p\in\mathit{pred}[x]$ \SuchThat $\mathit{winner}[p]=\PU$}{
       $\mathit{count}[p] \gets \mathit{count}[p] - 1$\label{ln:cntupdate}\;
       \lIf{$\mathit{owner}[p] = w \lor (\mathit{count}[p] = 0\land \mathit{frozen}[p])$\label{ln:cntchk2}}{
         \setwinner{$p, w$}\label{ln:propagate}%
       }
     }
 }
\end{algorithm}

  \LinesNumbered
\begin{algorithm}[tb]
  \caption{On-the-fly realizability check with Mealy semantics (for Moore semantics,
    swap the order of $\CI$ and $\CO$ on the first line).     \vspace*{-10mm}
    \label{algo:realizability}}

  \Fn{\realizability{$\varphi\in\LTLf(\CI\uplus\CO)$}}{
    configure the MTBDD library to put variables in $\CI$ before those in $\CO$\;
    $\mathit{init}\gets\term(\varphi,\bot)$; \newvertex{$\mathit{init},\PI$}\label{ln:init}\;
    $\CV\gets \{\mathit{init}\}$\tcp*[l]{nodes created as game vertices}
    $\CQ\gets \emptyset$\label{ln:bfs0}\tcp*[l]{$\LTLf$ formulas processed by main loop on line~\ref{ln:bfs1}}
    \Fn{\declarevertex{$r\in\MTBDD(\CI\uplus\CO,\LTLf(\CI\uplus\CO)\times\BB)$}}{
      $(p,\ell,h)\gets r$;
      \lIfElse{$p = \infty\lor p\in \CI$}{$\mathit{own}\gets\PI$}{$\mathit{own}\gets\PO$}
      \newvertex{$r, \mathit{own}$};
      $\CV\gets\CV\cup \{r\}$;
      $\mathit{to\_encode}\gets\mathit{to\_encode}\cup \{r\}$\;
    }
    $\mathit{todo}\gets\{\varphi\}$\label{ln:bfs1b}\;
    \While{$\mathit{todo}\ne\emptyset\land \mathit{winner}[\mathit{init}] = \PU$\label{ln:bfs1}}{
      $\alpha\gets \mathit{todo}.\mathsf{pop\_any}()$;
      $\CQ\gets\CQ\cup\{\alpha\}$\label{ln:bfs2}\;
      \textcolor{gray}{[optional: add one-step (un)realizability check here, see Sec.~\ref{sec:onestep}]}\;
      $a\gets \term(\alpha, \bot)$; $m\gets \tr(\alpha)$\label{ln:dfs3}\;
      \If(\tcp*[h]{$m$ has already been encoded}){$m\in \CV$}{
        \newedge{$a,m$};\,\freezevertex{$a$};
        \Continue to line~\ref{ln:bfs1}\;
       }
      $\mathit{to\_encode}\gets\emptyset$; $\mathit{leaves}\gets\emptyset$\;
      $\declarevertex{m}$; \newedge{$a,m$}; \freezevertex{$a$}\;
      \lIf{$\mathit{winner}[a] \ne \PU$}{\Continue to line~\ref{ln:bfs1}\label{ln:deter1}}
      \While{$\mathit{to\_encode}\ne\emptyset$\label{ln:while1}}{
        $r \gets \mathit{to\_encode}.\mathsf{pop\_any}()$\;
        $(p,\ell,h)\gets r$\;
        \uIf(\tcp*[h]{this is a terminal labeled by $\ell$}){$p = \infty$}{
          $(\beta,b)\gets\ell$\;
          \lIf{b}{\setwinner{$r, \PO$}\label{ln:O}}
          \lElseIf{$\beta=\ffalse$}{\setwinner{$r, \PI$}\label{ln:I}}
          \lElseIf{$\beta\not\in\CQ$}{$\mathit{leaves}\gets\mathit{leaves}\cup\{\beta\}$\label{ln:S}}
        }\Else{
          \lIf{$\ell\not\in\CV$}{\declarevertex{$\ell$}}
          \lIf{$h\not\in\CV$}{\declarevertex{$h$}}
          \newedge{$r,\ell$};
          \newedge{$r,h$};
          \freezevertex{$r$}\;
        }
        \lIf{$\mathit{winner}[a] \ne \PU$}{\Continue to line~\ref{ln:bfs1}\label{ln:deter2}\label{ln:while2}}
      }
      $\mathit{todo}\gets\mathit{todo}\cup\mathit{leaves}$\label{ln:bfs3}\;
    }
    \Return $\mathit{winner}[\mathit{init}]=\PO$\label{ln:lastret}\;
  }
\end{algorithm}
  \LinesNotNumbered

  Algorithm~\ref{algo:realizability} is the second part.  It shows how to build $\CG_\varphi$ incrementally.  It translates the states $\alpha$ of the corresponding MTDFA one at a time, and uses the functions of Algorithm~\ref{algo:gamesolving} to turn each node of $\tr(\alpha)$ into a vertex of the game.  Since the functions of Algorithm~\ref{algo:gamesolving} update the winning status of the states as soon as possible, Algorithm~\ref{algo:realizability} can use that to cut parts of the exploration.

  Instead of using $\Delta(\varphi)=\tr(\varphi)$ as initial vertex of
  the game, as in Theorem~\ref{th:encode}, we consider
  $\mathit{init}=\term(\varphi,\bot)$ as initial vertex
  (line~\ref{ln:init}): this makes no theoretical difference, since
  $\term(\varphi,\bot)$ has $\tr(\varphi)$ as unique successor.
  Lines \ref{ln:bfs0},\ref{ln:bfs1b}--\ref{ln:bfs2},\ref{ln:dfs3}, and
  \ref{ln:bfs3} implements the exploration of all the $\LTLf$ formulas
  $\alpha$ that would label the states of the MTDFA for $\varphi$ (as
  needed to implement Theorem~\ref{th:translation}).  The actual order
  in which formulas are removed from $\mathit{todo}$ on
  line~\ref{ln:bfs2} is free.  (We found out that handling
  $\mathit{todo}$ as a queue to implement a BFS exploration worked
  marginally better than using it as a stack to do a DFS-like
  exploration, so we use a BFS in practice.)\looseness=-1

  Each $\alpha$ is translated into an MTBDD $\tr(\alpha)$ representing
  its possible successors.  The constructed game should have one
  vertex per MTBDD node in $\tr(\alpha)$.  Those vertices are created
  in the inner \textbf{while} loop
  (lines~\ref{ln:while1}--\ref{ln:while2}).  Function~\declarevertex
  is used to assign the correct owner to each new node according to
  its decision variable (as in Def.~\ref{def:realgame}) as well as
  adding those nodes to the $\mathit{to\_encode}$ set processed by
  this inner loop.  Terminal nodes are either marked as winning for
  one of the players (lines~\ref{ln:O}--\ref{ln:I}) or stored in
  $\mathit{leaves}$ (line~\ref{ln:S}).

  Since connecting game vertices may backpropagate their winning
  status, the encoding loop can terminate early whenever the vertex
  associated to $\term(\alpha,\bot)$ becomes determined
  (lines~\ref{ln:deter1} and \ref{ln:deter2}).  If that vertex is not
  determined, the $\mathit{leaves}$ of $\alpha$ are added to
  $\mathit{todo}$ (line~\ref{ln:bfs3}) for further exploration.

  The entire construction can also stop as soon as the initial vertex
  is determined (line~\ref{ln:bfs1}).
  %Keep in mind that we are only interested in knowing if player
  %\PO can force the game to reach its targets.
  %Propagating the information that player \PI may force the game to
  %reach \marginpar{Cf.~App.~\ref{app:backproplose}} vertex $\ffalse$
  %can help us terminate earlier, but
  However, if the algorithm terminates with
  $\mathit{winning[init]}=\PU$, it still means that \PO cannot reach
  its targets.  Therefore, as tested by line~\ref{ln:lastret}, formula
  $\varphi$ is realizable iff $\mathit{winning}[init]=\PO$ in the
  end.% of the algorithm.

\begin{theorem}\label{th:on-the-fly}
  Algorithm~\ref{algo:realizability} returns $\ttrue$ iff
  $\varphi$ is Mealy-realizable.
\end{theorem}

\section{Implementation and Evaluation}\label{sec:eval}

\vspace*{-1ex}
Our algorithms have been implemented in
Spot~\cite{duret.22.cav}, after extending its fork of
BuDDy~\cite{buddy.99.manual} to support MTBDDs.
\marginpar{Cf. App.~\ref{app:mtbddimpl}} The release of Spot~2.14
distributes two new command-line tools:
\ExtLink{https://spot.lre.epita.fr/ltlf2dfa.html}{\texttt{ltlf2dfa}}
and
\ExtLink{https://spot.lre.epita.fr/ltlfsynt.html}{\texttt{ltlfsynt}},
implementing translation from \LTLf{} to MTDFA, and solving \LTLf{}
synthesis.  We describe and evaluate \texttt{ltlfsynt} in the following.

\vspace*{-1.5ex}
\paragraph{Preprocessing}\label{sec:preprocessing}
Before executing Algorithm~\ref{algo:realizability}, we use a few
preprocessing techniques to simplify the
problem.  We remove variables that always have the same polarity in
the specification (a simplification used also by
% this is NOT a heuristic
Strix~\cite{sickert.21.strix}), and
we decompose the specifications into output-disjoint
sub-specifications that can be solved
independently~\cite{finkbeiner.21.nfm}.  A specification such as $\Psi_1\land\Psi_2$, from Example~\ref{ex:psi}, is not solved directly as demonstrated here, but split into two output-disjoint specifications $\Psi_1$ and $\Psi_2$ that are solved separately. Finally, we also simplify $\LTLf$ formulas using very simple rewriting
\marginpar{Cf.~App.~\ref{app:simplify}}
rules such as $\X(\alpha)\land\X(\beta)\leadsto\X(\alpha\land\beta)$
that reduce the number of MTBDD operations required during translation.

\vspace*{-1.5ex}
\paragraph{One-step (un)realizability checks}\label{sec:onestep}
An additional optimization consists in performing one-step
realizability and one-step unrealizability checks in
Algorithm~\ref{algo:realizability}.
\marginpar{Cf. App.~\ref{app:onestep}} The principle is to transform
the formula $\alpha$ into two smaller Boolean formulas $\alpha_r$ and
$\alpha_u$, such that if $\alpha_r$ is realizable it implies that
$\alpha$ is realizable, and if $\alpha_u$ is unrealizable it implies
that $\alpha$ is unrealizable~\cite[Theorems 2--3]{xiao.21.aaai}.
Those Boolean formulas can be translated to BDDs for which
realizability can be checked by quantification.  On success, it avoids
the translation of the larger formula $\alpha$.  The simple formula
$\Psi_1\land\Psi_2$ of our running example is actually one-step
realizable.

\vspace*{-1.5ex}
\paragraph{Synthesis}\label{sec:synthesis}
After deciding realizability, \texttt{ltlfsynt} is able to extract a
%\marginpar{Cf.~App.~\ref{app:synthesis}}
strategy from the solved game in the form of a Mealy machine, and
encode that into an And-Invert Graph (AIG)~\cite{biere.11.tr}: the
expected output of the Synthesis Competition for the $\LTLf$ synthesis
tracks.  The conversion from Mealy to AIG reuses prior
work~\cite{renkin.22.forte,renkin.23.fmsd} developed for Spot's
\textsf{LTL} (not $\LTLf$) synthesis tool.  We do not detail nor
evaluate these extra steps here due to lack of
space. % and confidence: \texttt{ltlfsynt} is
% currently the only tool producing AIG from $\LTLf$ synthesis, and we
% do not yet have tests to ensure the correctness of the produced AIG.

\vspace*{-1.5ex}
\paragraph{Evaluation}%\label{sec:eval}
We evaluated the task of deciding $\LTLf$ reachability over
specifications from the Synthesis Competition~\cite{jacobs.22.arxiv}.
We took all \texttt{tlsf-fin} specifications from
\ExtLink{https://github.com/SYNTCOMP/benchmarks/tree/447beee22384dba/tlsf-fin}{SyntComp's
  repository}, excluded some duplicate specifications as well as some
specifications that were too large to be solved by any tool, and
converted the specifications from TLSF v1.2~\cite{jacobs.23.tlsf12} to
$\LTLf$ using \texttt{syfco}~\cite{jacobs.23.tlsf12}.

We used BenchExec~3.22~\cite{beyer.19.sttt} to
track time and memory usage of each tool.  Tasks
were run on a Core i7-3770 with \emph{Turbo Boost} disabled, and
frequency scaled down to 1.6GHz to prevent CPU throttling.  The
computer has 4 physical cores and 16GB of memory.  BenchExec was
configured to run up to 3 tasks in parallel with a memory limit of 4GB
per task, and a time limit of 15 minutes.

\begin{figure}[tb]
  % Use cactus-crop.pdf if you prefer horizontal arrangement.
  \includegraphics[width=\textwidth]{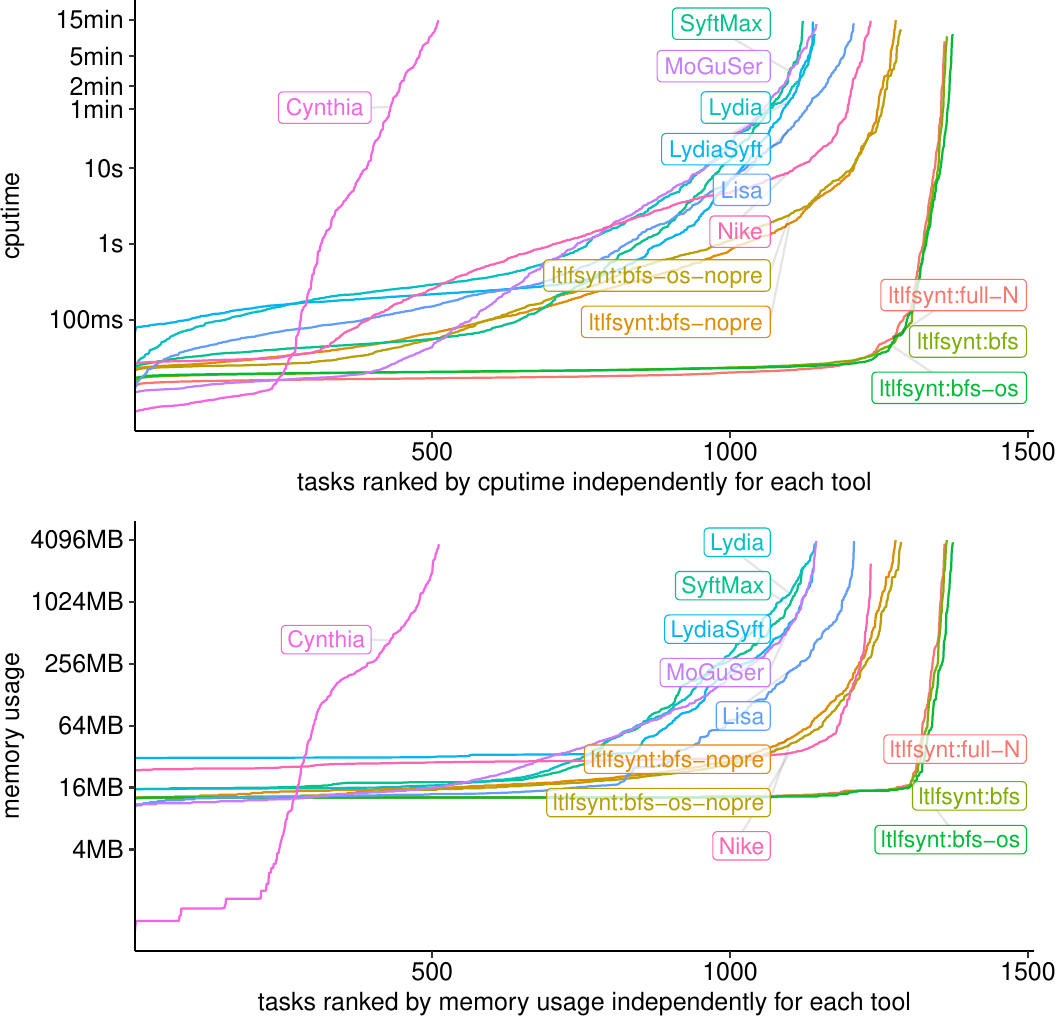}
         \vspace*{-7mm}
  \caption{Cactus plots comparing time and memory usage of different
    configurations.\label{fig:cactus}}
         \vspace*{-5mm}
\end{figure}

Figure~\ref{fig:cactus} compares five configurations of
\texttt{ltlfsynt} against seven other tools.
\marginpar{More in App.~\ref{app:benchmark}.}
We verified that all tools were in agreement.
Lydia~0.1.3~\cite{giacomo.21.icaps}, SyftMax~(or Syft
2.0)~\cite{zhu.22.ijcai} and LydiaSyft
0.1.0-alpha~\cite{favorito.25.tacas} are all using Mona to construct a
DFA by composition; they then solve the resulting
game symbolically after encoding it using BDDs.
Lisa~\cite{bansal.20.aaai} uses a hybrid compositional construction,
mixing explicit compositions (using Spot), with symbolic compositions
(using BuDDy), solving the game symbolically in the end.
Cynthia~0.1.0~\cite{degiacomo.22.ijcai},
Nike~0.1.0~\cite{favorito.23.rcra}, and MoGuSer~\cite{xiao.2024.vmcai}
all use an on-the-fly construction of a DFA game that they solve
via forward exploration with backpropagation, but they do not implement backpropagation in linear time, as we do.  Yet, the costly
part of synthesis is game generation, not solving.  Cynthia uses
SDDs~\cite{darwiche.11.ijcai} to compute successors and represent
states, while Nike and MoGuSer use SAT-based techniques to compute
successors and BDDs to represent states.
 Nike, Lisa, and LydiaSyft
were the top-3 contenders of the \LTLf{} track of SyntComp in 2023 and 2024.
%\looseness=-1

Configuration~\texttt{ltlfsynt:bfs-nopre} corresponds to
Algorithm~\ref{algo:realizability} were $\mathit{todo}$ is a queue: it
already solves more cases than all other tested tools.  Suffix
\texttt{-nopre} indicates that preprocessings of the specification
are disabled (this makes comparison fairer, since other tools
have no such preprocessings).  The version with preprocessings enabled
is simply called \texttt{ltlfsynt:bfs}.  Variants with
``\texttt{-os}'' adds the one-step (un)realizability checks that
LydiaSyft, Cynthia, and Nike also perform.\looseness=-1

We also include a configuration \texttt{ltlfsynt:full-N} that
corresponds to first translating the specification into a MTDFA using
Theorem~\ref{th:translation}, and then solving the game by linear
propagation.  The difference between \texttt{ltlfsynt:full} and
\texttt{ltlfsynt:bfs} shows the gain obtained with the on-the-fly
translation: although that look small in the cactus plot, it is
important in some
specifications.\marginpar{Tab.~\ref{tab:counter}--\ref{tab:counters}
  in App.~\ref{app:benchmark}.}

\vspace*{-1.5ex}
\paragraph{Data Availability Statement}
Implementation, supporting scripts, detailed analysis of this
benchmark, and additional examples are archived on
Zenodo~\cite{duret.25.zenodo}.

\vspace*{-1.5ex}
\section{Conclusion}
\vspace*{-1ex}
We have presented the implementation of \texttt{ltlfsynt}, and
evaluated it to be faster at deciding $\LTLf$ realizability than seven
existing tools, including the winners of SyntComp'24.  The implementation
uses a direct and efficient translation from $\LTLf$ to DFA represented by
MTBDDs, which can then be solved as a game played directly on the structure of
the MTBDDs.
The two constructions (translation and game
solving) are performed together on-the-fly, to allow early termination.

Although \texttt{ltlsynt} also
includes a preliminary implementation of $\LTLf$ synthesis of
And-Inverter graphs, we leave it as future work
to document it and ensure its correctness.%\looseness=-1

Finally, the need for solving a reachability game while it is
discovered also occurs in other equivalent contexts such as HornSAT,
where linear algorithms that do not use ``counters'' and
``predecessors'' (unlike ours) have been
developed~\cite{liu.98.icalp}.  Using such algorithms might improve
our solution by saving memory.

\bibliographystyle{splncs04}
\bibliography{biblio}

\ifappendix
\newpage
\appendix

These appendices and the margin notes that point to them were
part of the submission for interested reviewers, but they have not
been peer-reviewed, and are not part of the CIAA'25 proceedings.

\section{MTBDD operations}\label{app:mtbdds}

This section details some of the MTBDD operations described in
section~\ref{sec:mtbdd}.  We believe those functions should appear
straightforward to any reader familiar with BDD implementations.
We show them for the sake of being comprehensive.

\subsection{Evaluating an MTBDD using an assignment}\label{app:mtbddeval}

For $m\in\MTBDD(\CP,\CS)$, and $w\in\BB^\CP$,
Algorithm~\ref{algo:eval} shows how to compute $m(w)$ by
descending the structure of $m$ according to $w$.

\begin{algorithm}[h]
  \caption{Evaluating an MTBDD using an assignment.\label{algo:eval}}

  \Fn{\eval{$m,\,w$}}{
    \Input{$m\in\MTBDD(\CP,\CS)$, $w\in\BB^\CP$}
    \Output{$m(w)\in \CS$}
    \BlankLine{}

    $(p,\ell,h) \gets m$\;
    \While{$p \ne \infty$}{
      \lIfElse{$w(v)$}{
        $(p,\ell,h) \gets h$
      }{
        $(p,\ell,h) \gets \ell$
      }
    }
    \Return{} $\ell$\;
  }
\end{algorithm}

\subsection{Binary and Unary Operations on MTBDDs}\label{app:mtbddapply}

Algorithm~\ref{algo:apply2} shows how the implementation of \apply follows
a classical recursive definition typically found in BDD
packages~\cite{bryant.86.tc,fujita.97.fmsd,andersen.99.lecturenotes}.  The function
\texttt{makebdd} is in charge of ensuring the \emph{reduced} property
of the MTBDD: for any triplet of the form $(p,r,r)$ where the
$\bddlow$ and $\bddhigh$ links are equal, \texttt{makebdd} returns $r$
to skip over the node.  For other triplets, \texttt{makebdd} will look
up and possibly update a global hash table to ensure that each triplet
is represented only once.  The hash table $H$ is used for memoization;
assuming lossless caching (i.e., no dropped entry on hash collision),
this ensures that the number of recursive calls performed is at most
$|m_1|\cdot|m_2|$.  Our implementation, as discussed in
Section~\ref{app:mtbddimpl}, uses a lossy cache, therefore the
complexity might be higher.

\begin{algorithm}[tp]
  \caption{Composing two MTBDDs by applying a binary operator to their terminals.}
  \label{algo:apply2}

  \SetKwFunction{apply}{apply2}
  \Fn{\apply{$m_1,\,m_2,\,\odot,\,H$}}{
    \Input{$m_1\in\MTBDD(\CP,\CS_1)$, $m_2\in\MTBDD(\CP,\CS_2)$, $\odot:\CS_1\times\CS_2\to\CS_3$,
    $H: \hashmap$}
    \Output{$m_1\odot m_2 \in\MTBDD(\CP,\CS_3)$}
    \BlankLine{}

    \If{$(m_1,m_2,\odot)\in H$}{
      \Return{} $H[(m_1,m_2,\odot)]$ \;
    }
    $(p_1,\ell_1,h_1) \gets m_1$\;
    $(p_2,\ell_2,h_2) \gets m_2$\;
    \uIf{$p_1 < p_2$}{
      $r \gets{}$ \mk{$p_1$, \apply{$\ell_1,\,m_2,\,\odot,\,H$}, \apply{$h_1,\,m_2,\,\odot,\,H$}}\;
    }
    \uElseIf{$p_2 < p_1$}{
      $r \gets{}$ \mk{$p_2$, \apply{$m_1,\,\ell_2,\,\odot,\,H$}, \apply{$m_1,\,h_2,\,\odot,\,H$}}\;
    }
    \uElseIf(\tcp*[h]{$p_1 = p_2$}){$p_1 < \infty$}{
      $r \gets{}$ \mk{$p_1$, \apply{$\ell_1,\,\ell_2,\,\odot,\,H$}, \apply{$h_1,\,h_2,\,\odot,\,H$}}\;
    }
    \Else(\tcp*[h]{$p_1=p_2=\infty$, we have terminals holding values $\ell_1$ and $\ell_2$}){
      $r \gets{}$ \mk{$\infty,\ell_1\odot\ell_2,\infty$}\;
    }
    $H[(m_1,m_2,\odot)] \gets r$\;
    \Return{} $r$\;
  }
\end{algorithm}
\begin{algorithm}[tbp]
  \caption{Gathering the leaves of an MTBDD can be done with a simple
    linear traversal of an MTBDD.\label{algo:leavesof}}

  \Fn{\leaves{$m$}}{
    \Input{$m\in\MTBDD(\CP,\CS)$}
    \Output{the subset of $\CS$ that appears on leaves of $m$}
    \BlankLine{}
    $\mathit{seen}\gets\{m\}$\;
    $\mathit{todo}\gets\{m\}$\;
    $\mathit{res}\gets \emptyset$\;
    \While{$\mathit{todo}\ne\emptyset$}{
      $m\gets \mathit{todo}.\mathsf{pop\_any}()$\;
      $(p,\ell,h)\gets m$\;
      \uIf(\tcp*[h]{We reached a leaf labeled by $\ell$}){$p=\infty$}{
        $\mathit{res}\gets \mathit{res}\cup\{\ell\}$\;
      }\Else{
        $\mathit{todo}\gets \mathit{todo}\cup(\{\ell,r\}\setminus\mathit{seen})$\;
        $\mathit{seen}\gets \mathit{seen}\cup\{\ell,r\}$\;
      }
    }
    \Return{} $\mathit{res}$\;
  }
\end{algorithm}

An \applyone function can be written along the same lines for
unary operators.

\subsection{Leaves of an MTBDD}\label{app:leaves}

Function \leaves{$m$}, shown by Algorithm~\ref{algo:leavesof} is a
straightforward way to collect the leaves that appear in an MTBDD $m$.

\subsection{Boolean Operations with Shortcuts}\label{app:apply2sc}

Algorithm~\ref{algo:apply2sc} shows how to implement Boolean operations
on MTBDDs with terminals in $\CS\cup\BB$, shortcutting the recursion
when one of the operands is a terminal labeled by a value in $\BB$.

\begin{algorithm}[t]
  \caption[Variant of \texttt{apply2}]{Variant of \apply that
    implements shortcuts when one of the argument is a Boolean leaf.}
  \label{algo:apply2sc}

  \Fn{\applysc{$m_1,\,m_2,\,\odot,\,H$}}{
    \Input{$m_1\in\MTBDD(\CP,\CS_1\cup\BB)$, $m_2\in\MTBDD(\CP,\CS_2\cup\BB)$, $\odot:\CS_1\times\CS_2\to\CS_3$,
    $H: \hashmap$}
    \Output{$m_1\odot m_2 \in\MTBDD(\CP,\CS_3)$}
    \BlankLine{}

    \If{$(m_1,m_2,\odot)\in H$}{
      \Return{} $H[(m_1,m_2,\odot)]$ \;
    }
    $(p_1,\ell_1,h_1) \gets m_1$\;
    $(p_2,\ell_2,h_2) \gets m_2$\;
    \If{$(p_1 = \infty \land \ell_1\in\BB) \lor (p_2 = \infty \land \ell_2\in\BB)$}{
      \Switch{$\odot$}{
        \Case{$\land$}{
           \lIf{$\bot\in\{\ell_1,\ell_2\}$}{\Return $(\infty,\bot,\infty)$}
           \lIf{$\ell_1=\top$}{\Return $m_2$}
           \lIf{$\ell_2=\top$}{\Return $m_1$}
          }
        \Case{$\lor$}{
           \lIf{$\top\in\{\ell_1,\ell_2\}$}{\Return $(\infty,\top,\infty)$}
           \lIf{$\ell_1=\bot$}{\Return $m_2$}
           \lIf{$\ell_2=\bot$}{\Return $m_1$}
          }
          \lCase{$\ldots$}{$\ldots$}
        }
      }

    \uIf{$p_1 < p_2$}{
      $r \gets{}$ \mk{$p_1$, \applysc{$\ell_1,\,m_2,\,\odot,\,H$}, \applysc{$h_1,\,m_2,\,\odot,\,H$}}\;
    }
    \uElseIf{$p_2 < p_1$}{
      $r \gets{}$ \mk{$p_2$, \applysc{$m_1,\,\ell_2,\,\odot,\,H$}, \applysc{$m_1,\,h_2,\,\odot,\,H$}}\;
    }
    \uElseIf(\tcp*[h]{$p_1 = p_2$}){$p_1 < \infty$}{
      $r \gets{}$ \mk{$p_1$, \applysc{$\ell_1,\,\ell_2,\,\odot,\,H$}, \applysc{$h_1,\,h_2,\,\odot,\,H$}}\;
    }
    \Else(\tcp*[h]{$p_1=p_2=\infty$, we have terminals holding values $\ell_1$ and $\ell_2$}){
      $r \gets{}$ \mk{$\infty,\ell_1\odot\ell_2,\infty$}\;
    }
    $H[(m_1,m_2,\odot)] \gets r$\;
    \Return{} $r$\;
  }
\end{algorithm}

\section{Boolean Operations on MTDFAs}\label{app:mtdfaops}

Although it is not necessary for the approach we presented, our
implementation supports all Boolean operations over MTDFAs.

Since $\Delta(q)\in\MTBDD(\CP,\CQ\times\BB)$ has terminals labeled by pairs
of the form $(q,b)\in\CQ\times\BB$, let us extend any Boolean operator
$\odot: \BB\times\BB\to\BB$ so that it can work on such pairs.  More
formally, for $(q_1,b_1)\in\CQ_1\times \BB$ and
$(q_2,b_2)\in\CQ_2\times \BB$ we define $(q_1,b_1)\odot(q_2,b_2)$ to
be equal to
$((q_1,q_2),(b_1\odot b_2)) \in ((Q_1\times Q_2)\times \BB)$.  Using
Algorithm~\ref{algo:apply2} to apply $\odot$ elements of
$\MTBDD(\CP,\CQ\times\BB)$ gives us a very simple way to combine
MTDFAs, as shown by the following definition.

\begin{definition}[Composition of two MTDFAs]\label{def:mtdfa-compose}
  Let $\CA_1=\langle \CQ_1, \CP, \iota_1, \Delta_1\rangle$ and
  $\CA_2=\langle \CQ_2, \CP, \iota_2, \Delta_2\rangle$ be two MTDFAs over
  the same variables $\CP$, and let
  $\odot\in\{\land,\lor,\rightarrow,\leftrightarrow,...\}$ be any
  Boolean binary operator.

  Then, let $\CA_1\odot\CA_2$ denote the composition of $\CA_1$ and
  $\CA_2$ defined as the
  MTDFA
  $\langle \CQ_1\times \CQ_2, \CP, (\iota_1,\iota_2), \Delta\rangle$
  where for any $(q_1,q_2)\in \CQ_1\times \CQ_2$ we have
  $\Delta_3((q_1,q_2))=\Delta_1(q_1)\odot \Delta_2(q_2)$.
\end{definition}

\begin{property}
  With the notations from Definition~\ref{def:mtdfa-compose},
  $\lang(\CA_1\odot\CA_2) = \{ \sigma\in (\BB^\CP)^+ \mid (\sigma \in
  \lang(\CA_1)) \odot (\sigma \in \lang(\CA_2)) \}$.  In particular
  $\lang(\CA_1\land\CA_2)=\lang(\CA_1)\cap\lang(\CA_2)$ and
  $\lang(\CA_1\lor\CA_2)=\lang(\CA_1)\cup\lang(\CA_2)$.  If $\oplus$
  designates the \emph{exclusive or} operator, testing the equivalence
  of two automata $\lang(\CA_1)=\lang(\CA_2)$ amounts to testing
  whether $\lang(\CA_1\oplus\CA_2)=\emptyset$.
\end{property}

The complementation of an MTDFA (with respect to $(\BB^\CP)^+$ not
$(\BB^\CP)^\star$) can be defined using the unary Boolean negation
similarly.

Such compositional operations are at the heart of the compositional \LTLf{} translations used by Lisa~\cite{bansal.20.aaai}, Lydia~\cite{giacomo.21.icaps} and LydiaSyft~\cite{favorito.25.tacas}.  This is efficient as it allows minimizing intermediate automata before combining them.  Our translator tool \ExtLink{https://spot.lre.epita.fr/ltlf2dfa.html}{\texttt{ltlf2dfa}} uses such a compositional approach by default.   For \LTLf{} synthesis, our tool \ExtLink{https://spot.lre.epita.fr/ltlfsynt.html}{\texttt{ltlfsynt}} also
has the option to build the automaton by composition, but this is not enabled by default: using an on-the-fly construction as presented in Algorithm~\ref{algo:realizability} is more efficient.  We refer the reader to the artifact~\cite{duret.25.zenodo} for benchmark comparisons involving our own implementation of the compositional approach.

\section{Simplified MTDFA}\label{app:simplified}

\begin{figure}[tb]
  \begin{tikzpicture}[scale=.74]
    \node[root,left] at (0,-0.5) (r0) {$\Psi_1 \land ((\G\F o_2){\liff}(\F i_0))$};
    \node[left] at (r0.west) (init) {$\iota=$};
    \node[root,left] at (0,-2) (r2) {$\ffalse$};
    \node[root,left] at (0,-3.5) (r3) {$\Psi_1 \land (\G\F o_2)$};
    \node[inode] at (1, -0.5) (i00) {$i_0$};
    \node[inode] at (1, -3.5) (i01) {$i_0$};
    \node[inode] at (2.2, -0.5) (i20) {$i_2$};
    \node[inode] at (2.2, -3) (i10) {$i_1$};
    \node[inode] at (2.2, -4) (i21) {$i_2$};
    \node[inode] at (3.4, -0) (o10) {$o_1$};
    \node[inode] at (3.4, -1) (o11) {$o_1$};
    \node[inode] at (3.4, -3) (o12) {$o_1$};
    \node[inode] at (3.4, -4) (o13) {$o_1$};
    \node[inode] at (4.6, -0) (o20) {$o_2$};
    \node[inode] at (4.6, -4) (o21) {$o_2$};
    \node[termn,right] at (5.6,0) (t0) {$\Psi_1 \land ((\G\F o_2){\liff}(\F i_0))$};
    \node[terma,right] at (5.6,-1) (t1) {$\Psi_1 \land ((\G\F o_2){\liff}(\F i_0))$};
    \node[termn,right] at (5.6,-2) (t2) {$\ffalse$};
    \node[terma,right] at (5.6,-3) (t3) {$\Psi_1 \land (\G\F o_2)$};
    \node[termn,right] at (5.6,-4) (t4) {$\Psi_1 \land (\G\F o_2)$};
    \draw[high] (i00) -> (i10);
    \draw[low]  (i00) -> (i20);
    \draw[high] (i01) -> (i10);
    \draw[low]  (i01) -> (i21);
    \draw[high] (i20) -> (o10);
    \draw[low]  (i20) -> (o11);
    \draw[high] (i10) -> (o13);
    \draw[low]  (i10) -> (o12);
    \draw[high] (i21) -> (o13);
    \draw[low]  (i21) -> (o12);
    \draw[high] (o10) -> (o20);
    \draw[low]  (o10) -> (t2);
    \draw[high] (o11) -> (t2);
    \draw[low]  (o11) -> (o20);
    \draw[high] (o12) -> (t2);
    \draw[low]  (o12) -> (o21);
    \draw[high] (o13) -> (o21);
    \draw[low]  (o13) -> (t2);
    \draw[high] (o20) -> (t0);
    \draw[low]  (o20) -> (t1.west);
    \draw[high] (o21) -> (t3.west);
    \draw[low]  (o21) -> (t4);
    \draw[rlink] (r0.east) -> (i00);
    \draw[rlink] (r2.east) -> (t2);
    \draw[rlink] (r3.east) -> (i01);
  \end{tikzpicture}
  \caption[Simplified MTDFA]{The MTDFA from Figure~\ref{fig:mtdfa}, simplified by merging all states that have an identical MTBDD successor and adjusting the terminals.\label{fig:mtdfa-trans}}
 % \end{figure}
 % keep these two figures together.
 %\begin{figure}[b]
  \quad~
  \begin{tikzpicture}[automaton,every state/.style={rectangle,rounded corners=2ex}]
    \node[initial,state] (i) {$\Psi_1 \land ((\G\F o_2){\liff}(\F i_0))$};
    \node[state,right=of i] (o) {$\Psi_1 \land (\G\F o_2)$};
    \draw[double>] (i) to[out=120,in=70,looseness=5] node[above]{
      $\lnot i_0 \land (i_2 \liff o_1) \land o_2$} (i);
    \draw[->] (i) to[in=-120,out=-70,looseness=5] node[below]{
      $\lnot i_0 \land (i_2 \liff o_1) \land \lnot o_2$} (i);
    \draw[double>] (i) to[out=10,in=166] node[above]{
      $i_0 \land (i_1 \liff o_1) \land o_2$} (o);
    \draw[->] (i) to[out=-10,in=-166] node[below]{
      $i_0 \land (i_1 \liff o_1) \land \lnot o_2$} (o);
    \draw[double>] (o) to[out=120,in=70,looseness=5] node[above]{
      $\bigl((\lnot i_0 \land (i_2 \liff o_1))\lor(\mathrlap{i_0 \land (i_1 \liff o_1))\bigr) \land o_2}$} (o);
    \draw[->] (o) to[in=-120,out=-70,looseness=5] node[below]{
      $\bigl((\lnot i_0 \land (i_2 \liff o_1))\lor(\mathrlap{i_0 \land (i_1 \liff o_1))\bigr) \land \lnot o_2}$} (o);
  \end{tikzpicture}
  \caption[TDFA]{Transition-based DFA interpretation of the MTDFA of
    Figure~\ref{fig:mtdfa-trans}.  A word $\sigma\in(\CP^\BB)^+$ is
    accepted if there is a run of the automaton such that each
    assignment $\sigma(i)$ is compatible with the Boolean formula
    labeling the transition, and if the last transition visited was
    accepting (double line).  The sink state corresponding to $\ffalse$ has
    been trimmed for clarity.\label{fig:tdfa}}
\end{figure}

Figure~\ref{fig:mtdfa-trans} shows a simplified version of the MTDFA
from Figure~\ref{fig:mtdfa} that can be obtained by any one of two
optimizations discussed in Section~\ref{sec:trans-optims}:
\begin{itemize}
\item merge states with identical MTBDD representations, or
\item apply the $(\G\beta)\land \beta\leadsto\G\beta$ simplification during constriction.
\end{itemize}
The second optimization is faster, as it does not require
computing $\tr(q)$ on some state $q$ only to later find that the
result is identical to some previous $\tr(q')$.

This simplified automaton may also help understand the
``transition-based'' nature of those MTDFAs.  Here we have pairs of
terminal with identical formula labels, but different acceptance:
words are allowed to finish on one, but not the other.  If they
continue, they continue from the state specified by the formula.
Figure~\ref{fig:tdfa} shows an equivalent ``transition-based DFA'' using notations that should be more readable by readers familiar with finite
automata.

\section{Try it Online!}\label{app:tryonline}

The \ExtLink{https://spot-sandbox.lre.epita.fr/}{Spot Sandbox} website
offers online access to the development version of Spot (which
includes the work described here) via Jupyter
notebooks~\cite{jupyter.16.elpub} or shell terminals.

In order to try the
\ExtLink{https://spot.lre.epita.fr/ltlf2dfa.html}{\texttt{ltlf2dfa}}
and
\ExtLink{https://spot.lre.epita.fr/ltlfsynt.html}{\texttt{ltlfsynt}}
command-line tools, simply connect to
\ExtLink{https://spot-sandbox.lre.epita.fr/}{Spot Sandbox}, hit the
``New'' button, and start a ``Terminal''.

The example directory contains two Jupyter notebooks directly related
to this submission:
\begin{itemize}
\item \ExtLink{https://spot-sandbox.lrde.epita.fr/notebooks/examples (read only)/backprop.ipynb}{backprop.ipynb} illustrates Algorithm~\ref{algo:gamesolving}.  There, players \PO and \textsc{1}
are called \texttt{True} and \texttt{False} respectively.
\item \ExtLink{https://spot-sandbox.lrde.epita.fr/notebooks/examples (read only)/ltlf2dfa.ipynb}{ltlf2dfa.ipynb} illustrates the translation of Section~\ref{sec:translate} with the optimizations
  discussed in Section~\ref{sec:trans-optims} (page~\ref{sec:trans-optims}), the MTDFA operations mentioned in Appendix~\ref{app:mtdfaops}, and some other game solving techniques not discussed here.
\end{itemize}

An HTML version of these two notebooks can also be found in directory
\texttt{more-examples/} of the associated
artifact~\cite{duret.25.zenodo}.

\section{Backpropagation of Losing Vertices}\label{app:backproplose}

The example of Figure~\ref{fig:game} does not make it very clear how
marking $\ffalse$ as a losing vertex (i.e., winning for \PI) may
improve the on-the-fly game solving: it does not help in that example.

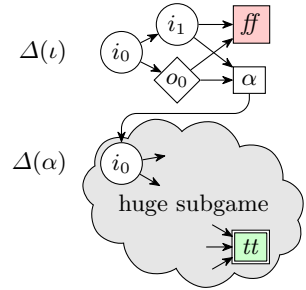
\begin{figure}[t]
  \centering
  \begin{tikzpicture}[scale=.74]
    \node[inode] at (0, 0) (i0) {$i_0$};
    \node[left=1em] at (i0.west) {$\Delta(\iota)$};
    \node[inode] at (1, 0.5) (i1) {$i_1$};
    \node[control] at (1, -0.5) (o0) {$o_0$};
    \node[lose,termn,right] at (2,0.5) (t0) {$\ffalse$};
    \node[termn,right] at (2,-0.5) (t1) {$\alpha$};
    \draw[->] (i0) -- (i1);
    \draw[->] (i0) -- (o0);
    \draw[->] (i1) -- (t0);
    \draw[->] (i1) -- (t1);
    \draw[->] (o0) -- (t0);
    \draw[->] (o0) -- (t1);
    \node [rotate=-15,cloud, draw,cloud puffs=11,cloud puff arc=120, aspect=2, inner ysep=1ex, fill=gray!20,minimum width=3cm,minimum height=2.2cm] at (1.3,-2.7) {};
    \node[inode,fill=white] at (0, -2) (i01) {$i_0$};
    \draw[->] (i01) -- ++(10:8mm);
    \draw[->] (i01) -- ++(-30:8mm);
   \draw[->,rounded corners=2mm] (t1) |- ($(t1)!.4!(i01)$) -| (i01);
    \node[win,terma,right] at (2, -3.5) (t2) {$\ttrue$};
    \draw[<-] (t2) -- ++(150:8mm);
    \draw[<-] (t2) -- ++(180:8mm);
    \draw[<-] (t2) -- ++(-150:8mm);
    \node[xshift=1mm] at ($(i01)!.5!(t2)$) {huge subgame};
    \node[left=1em] at (i01.west) {$\Delta(\alpha)$};
  \end{tikzpicture}
  \caption{If this game is created on-the-fly, it is useful to mark
    the $\ffalse$ terminal as losing.  When $\Delta(\iota)$ is encoded
    into a game, the fact that $\ffalse$ is winning for player \PI
    will cause the two round vertices above it to be immediately marked
    as winning as well.  Now, since the initial vertex is known to be
    winning for \PI, hence losing for \PO, the exploration may stop
    without having to encode $\Delta(\alpha)$.\label{fig:ffpropag}}
\end{figure}

Figure~\ref{fig:ffpropag} shows a scenario where marking states as
losing and propagating this information is useful to avoid some
unnecessary exploration of a large part of the automaton.
Algorithm~\ref{algo:realizability}, described in
Section~\ref{sec:realizability}, translates one state of the MTDFA at
a time, starting from $\iota$, and encodes that state into a game by
calling \newvertex, \newedge, etc.  In the example of
Figure~\ref{fig:ffpropag}, after the MTBDD for $\Delta(\iota)$ has
been encoded (the top five nodes of Figure~\ref{fig:ffpropag}), the
initial node will be marked as winning for \PI already (because \PI
can select the appropriate value of $i_0$ and $i_1$ to reach
$\ffalse$), therefore, the algorithm can stop immediately.  Had we
decided to backpropagate only the states winning for player $\PO$, the
algorithm would have to continue encoding $\Delta(\alpha)$ into
the game and probably many other states reachable from there.  At the
end of the backpropagation, the initial node would still be
undetermined, and we would also conclude that \PO cannot win.

Such an interruption of the on-the-fly exploration is used, does not only
occur when the initial state is determined.for the initial state, but at every  search:
if during the encoding of $\Delta(\alpha)$ we find that the winning status of
the root note of $\Delta(\alpha)$ is determined (line~\ref{ln:deter2}
of Algorithm~\ref{algo:realizability}), then it is unnecessary to
explore the rejecting leaves of $\Delta(\alpha)$.

\section{Implementation Details: MTBDDs in BuDDy}\label{app:mtbddimpl}

BuDDy~\cite{buddy.99.manual} is a BDD library created by J{\o}rn
Lind-Nielsen for his Ph.D. project.  Maintenance was passed to someone
else in 2004.  The Spot developer has contributed a few changes and
fixes to the ``original'' project, but it soon became apparent that
some of the changes motivated by Spot's needs could not be merged
upstream (e.g., because they would break other projects for the sake
of efficiency).  Nowadays, Spot is distributed with its own fork of
BuDDy that includes several extra functions, a more compact
representation of the BDD nodes (16 bytes par node instead of 20), a
``derecursived'' implementation of the most common BDD operations.
Moving away from BuDDy, to another BDD library would be very
challenging.  Therefore, for this work, we modified BuDDy to add
support for MTBDDs with \texttt{int}-valued terminals (our MTBDD
implementation knows nothing about $\LTLf$).

Our implementation differs from Mona's MTBDDs or CUDD's ADDs in
several ways.  First, BuDDy is designed around a global unicity table,
which stores reference counted BDDs.  There is no notion of ``BDD
manager'' as in Mona or CUDD that allows building independent BDDs.
We introduced support for MTBDD directly into this table, by reserving
the highest possible variable number to indicate a terminal (storing
the terminal's value in the $\bddlow$ link, as suggested by our
notation in this paper), and adding an extra \texttt{if} in the garbage
collector so it correctly deals with those nodes.  This change allows to
mix MTBDD terminals with regular BDD terminals (false and true).
Existing BDD function wills work as they have always done when a BDD
does not use the new terminals.  If multi-terminals are used, a new
set of functions should be used.

In CUDD's ADD implementation, the set of operations that can be passed
to the equivalent of the \apply function (see
Algorithm~\ref{algo:apply2}) is restricted to a fixed set of algebraic
operations that have well defined semantics.  In Mona and in our
implementation, the user may pass an arbitrary function in order to
interpret the terminals (which can only store an integer) and combine
them.  For instance, to implement the presented algorithm where
terminal are supposed to be labeled by pairs
$(\alpha,b)\in\LTLf(\CP)\times\BB$, we store $b$ in the lower bit of
the terminal's value, and use the other bits as an index in an array
that stores $\alpha$.  If we create a new formula while combining two
terminals, we add the new formula to that array, and build the value
of the newly formed terminal from the corresponding index in that
array.

One issue with implementing MTBDD operations is how to implement the
operation cache (the $H$ argument of Algorithm~\ref{algo:apply2}) when
the function to apply on the leaves is supplied by the user.  Since
the supplied function may depend on global variables, it is important
that this operation cache can be reset by the user.
%(Algorithms~\ref{algo:solveone} and~\ref{algo:solvetwo} show examples
%where the cache is reset between each iteration of the outer loop.)

We implement those user-controlled operation caches using lossy hash
tables similar to what are used internally by BuDDy for classical BDD
operations.  Algorithm~\ref{algo:apply2}, the line
$H[(m_1,m_2,\odot)] \gets r$ that saves the result of the last
operation may actually erase the result of a previous operation that
would have been hashed to the same index.  Therefore, the efficiency
of our MTBDD algorithms will depend on how many collisions they
generate, and this in turn depends on the size allocated for this hash
table: ideally $H$ should have a size of the same order as the number
of BDD nodes used in the MTBDD resulting from the operation.  We use
two empirical heuristics to estimate a size for $H$.  For unary
operations on MTDFAs, we set $|H|=|\CP|\cdot|\CQ|/2$, and for binary
operations on MTDFAs (e.g., Def.~\ref{def:mtdfa-compose}), we set
$|H|=|\CP_1\cup\CP_2|\cdot|\CQ_1|\cdot|\CQ_2|/4$.  For operations
performed during the translation of $\LTLf$ formulas to MTDFAs
(Th.~\ref{th:translation}), we use a hash table that is $20\%$ of the
total number of nodes allocated by BuDDy, but we share it for all
MTBDD operations performed during the translation.

Mona handles those caches differently: it also estimates
an initial size for those caches (with different
formulas~\cite{klarlund.96.tr}), but by default it will handle any
collision by chaining, growing an overflow table to store collisions
as needed.  This difference probably contributes to the additional
``out-of-memory'' errors that Mona-based tools tend to show in our
benchmarks.

\section{Simple Rewriting Rules}\label{app:simplify}

We use a specification decomposition technique based on~\cite{finkbeiner.21.nfm}.
We try to rewrite the input specification
$\varphi$ into a conjunction $\varphi=\bigwedge_i \varphi_i$, where
each $\varphi_i$ uses non-overlapping sets of outputs.  Formula
$\Psi=\Psi_1\land\Psi_2$ from Example~\ref{ex:psi} is already in this
form.  However, in general, the specification may be more complex,
like $\G(\xi_0)\limplies \bigwedge_i\xi_i$.  In such a case, we
rewrite the formula as $\bigwedge_i(\G(\xi_0)\limplies \xi_i)$ before
partitioning the terms of this conjunction into groups that use
overlapping sets of output variables.  Such a rewriting, necessary to an
effective decomposition, may introduce a lot of redundancy in the
formula (in this example $\G(\xi_0)$ is duplicated several times).

For this reason, we apply simple language-preserving rewritings on
$\LTLf{}$ formulas before attempting to translate them into an MTDFA.
These rewritings undo some of the changes that had to be done earlier
to look for possible decompositions.  They are also performed when
decomposition is disabled.  More generally, the goal is to reduce the
number of temporal operators, in order to reduce the number of MTBDD
operations that need to be performed.

\begin{align}
  (\alpha\limplies \beta)\land (\alpha\limplies \gamma) &\leadsto \alpha\limplies (\beta\land \gamma) \\
  (\alpha\limplies \beta)\lor (\gamma\limplies \delta) &\leadsto (\lnot \alpha)\lor \beta\lor(\lnot \gamma) \lor \delta \label{eq:or}\\
  \bigwedge_i\G(\alpha_i) \land \bigwedge_j\G\F(\beta_j)&\leadsto \G(\bigwedge_i\alpha_i\land \F(\bigwedge_j \beta_j))\\
  \bigvee_i\F(\alpha_i) \lor \bigvee_j\F\G(\beta_j)&\leadsto \F(\bigvee_i\alpha_i\lor \G(\bigvee_j \beta_j))\\
  \X\alpha\land \X\beta &\leadsto \X(\alpha\land\beta) \\
  \X\alpha\lor \X\beta &\leadsto \X(\alpha\lor\beta) \\
  \StrongX\alpha\land \StrongX\beta &\leadsto \StrongX(\alpha\land\beta) \\
  \StrongX\alpha\lor \StrongX\beta &\leadsto \StrongX(\alpha\lor\beta) \\
  \G\F(\alpha) &\leadsto \G\F(\alpha_r) \label{eq:gf1}\\
  \F\G(\alpha) &\leadsto \G\F(\alpha_r) \label{eq:fg1}
\end{align}

Equation \eqref{eq:or} is the only equation that does not reduce the
number of operators.  However, our implementation automatically removes
duplicate operands for $n$-ary operators such as $\land$ or $\lor$, so
this is more likely to occur after this rewriting.

In \LTLf{}, formulas $\G\F(\alpha)$ and $\F\G(\alpha)$ are equivalent,
and specify that $\alpha$ should hold on the last position of the
word.  Therefore, in \eqref{eq:gf1}--\eqref{eq:fg1}, any temporal
operators in $\alpha$ can be removed using the same rules as in
Theorem~\ref{th:onestepsat} in Appendix~\ref{app:onestep}.

\section{One-step (Un)Realizability Checks}\label{app:onestep}

To test if an \LTLf formula $\varphi$ is realizable or unrealizable in
one-step, we can rewrite the formula into Boolean formulas $\varphi_r$
or $\varphi_u$ using one of the following theorems that follow from
the \LTLf semantics.

Then testing whether a Boolean formula is (un)realizable can be
achieved by representing that formula as a BDD, and then removing
input/output variables by universal/existential quantification, in the
order required by the selected semantics (Moore or Mealy).

\begin{theorem}[One-step realizability~{\cite[Th.~2]{xiao.21.aaai}}]\label{th:onestepsat}
  For $\varphi\in\LTLf(\CP)$, define $\varphi_r$
  inductively using the following rules:
  \begin{align*}
    \ffalse_r &= \ffalse  & (\StrongX \alpha)_r &= \ffalse & (\G \alpha)_r &= \alpha_r & (\alpha \R \beta)_r &= \beta_r \\
    \ttrue_r &= \ttrue & (\X \alpha)_r &= \ttrue & (\F \alpha)_r &= \alpha_r & (\alpha \U \beta)_r &= \beta_r \\
  p_r &= p \mathrlap{\quad\text{for~}p\in\CP} &&& (\lnot \alpha)_r &= \lnot(\alpha_r)
    & (\alpha\odot\beta)_r &= \alpha_r \odot \beta_r
  \end{align*}
  Where $\odot\in\{\land,\lor,\limplies,\liff,\lxor\}$.

  If the Boolean formula $\varphi_r$ is realizable, then $\varphi$ is realizable too.
\end{theorem}

\begin{theorem}[One-step unrealizability~{\cite[Th.~3]{xiao.21.aaai}}]
  Consider a formula $\varphi\in\LTLf(\CP)$.  To simplify the
  definition, we assume $\varphi$ to be in \emph{negative normal form}
  (i.e., negations have been pushed down the syntactic tree, and may
  only occur in front of variables, and operators $\limplies$,
  $\liff$, $\lxor$ have been rewritten away).  We define $\varphi_u$
  inductively as follows:
  \begin{align*}
    \ffalse_u &= \ffalse  & (\StrongX \alpha)_u &= \ttrue & (\G \alpha)_u &= \alpha_u & (\alpha \R \beta)_u &= \alpha_u\land \beta_u & (\alpha\land\beta)_u &= \alpha_u \land \beta_u \\
    \ttrue_u &= \ttrue & (\X \alpha)_u &= \ttrue & (\F \alpha)_u &= \alpha_u & (\alpha \U \beta)_u &= \alpha_u\lor \beta_u & (\alpha\lor\beta)_u &= \alpha_u \lor \beta_u
  \end{align*}
  For any variable $p\in\CP$, we have $p_u=p$ and $(\lnot p)_u=\lnot p$.

  If the Boolean formula $\varphi_u$ is not realizable, then $\varphi$ is not realizable.
\end{theorem}

\section{More Benchmark Results}\label{app:benchmark}

The SyntComp benchmarks contain specifications that can be partitioned
in three groups:
\begin{description}
\item[game] Those specifications describe two-players games. They have
  three subfamilies~\cite{tabajara.19.ijcai}: \emph{single
    counter}, \emph{double counters}, and \emph{nim}.
\item[pattern] Those specifications are scalable patterns built
  either from nesting $\U$ operators, or by making conjunctions of terms such
  as $\G(v_i)$ or $\F(v_j)$.~\cite{xiao.21.aaai}
\item[random] Those specifications are random conjunctions of $\LTLf$
  specifications~\cite{zhu.17.ijcai}.
\end{description}

Of these three sets, the \emph{games} are the most challenging to
solve.  The \emph{patterns} use each variable only once, so they can
all be reduced to $\ttrue$ or $\ffalse$ by the preprocessing technique
discussed in Section~\ref{sec:preprocessing}, or by the one-step
(un)realizability checks.  Since \emph{random} specifications are
built as a conjunction of subspecifications that often have
nonintersecting variable sets, they can very often be decomposed into
output-disjoint specifications that can be solved
separately~\cite{finkbeiner.21.nfm}.

Table~\ref{tab:status} shows how the different tools succeed in
these different benchmarks.

\begin{table}
  \caption{Count of different tool outputs grouped by different types of benchmarks.
    Status \emph{false} and \emph{true} indicate how many times a tool successfully managed
    to decide unrealizability (\emph{false}) or realizability (\emph{true}).
    The other status are error conditions: TIMEOUT (over 15 minutes), OOMEM (over 4GB), ABORT (aborted), SEGV (segmentation violation).  The later
    two errors are likely caused by some incorrect handling of out-of-memory conditions.\label{tab:status}}
  \centering
  \adjustbox{totalheight=\textheight-7\baselineskip}{
\begin{tabular}{ll>{}c>{}cccccc}
\toprule
& & \multicolumn{6}{c}{status} & \multicolumn{1}{c}{} \\ \cmidrule(lr){3-8}
type & tool & false & true & TIMEOUT & OOMEM & ABORT & SEGV & \multicolumn{1}{c}{all} \\ 
\midrule
game & ltlfsynt:full-N  & \cellcolor{white}{$\phantom{00}40$} & \cellcolor{white}{$\phantom{0}45$} & $\phantom{0}19$ & $\phantom{0}16$ &  &  & $\phantom{0}120$ \\
 & ltlfsynt:bfs-nopre  & \cellcolor{white}{$\phantom{00}40$} & \cellcolor{white}{$\phantom{0}47$} & $\phantom{00}4$ & $\phantom{0}29$ &  &  & $\phantom{0}120$ \\
 & ltlfsynt:bfs-os-nopre  & \cellcolor{white}{$\phantom{00}40$} & \cellcolor{yellow}{$\phantom{0}57$} & $\phantom{00}3$ & $\phantom{0}20$ &  &  & $\phantom{0}120$ \\
 & ltlfsynt:bfs  & \cellcolor{yellow}{$\phantom{00}42$} & \cellcolor{white}{$\phantom{0}47$} & $\phantom{00}4$ & $\phantom{0}27$ &  &  & $\phantom{0}120$ \\
 & ltlfsynt:bfs-os  & \cellcolor{yellow}{$\phantom{00}42$} & \cellcolor{yellow}{$\phantom{0}57$} & $\phantom{00}3$ & $\phantom{0}18$ &  &  & $\phantom{0}120$ \\
 & SyftMax  & \cellcolor{white}{$\phantom{000}7$} & \cellcolor{white}{$\phantom{0}29$} & $\phantom{0}23$ & $\phantom{0}45$ & $1$ & $15$ & $\phantom{0}120$ \\
 & Lydia  & \cellcolor{white}{$\phantom{000}7$} & \cellcolor{white}{$\phantom{0}28$} & $\phantom{0}24$ & $\phantom{0}45$ & $1$ & $15$ & $\phantom{0}120$ \\
 & LydiaSyft  & \cellcolor{white}{$\phantom{000}7$} & \cellcolor{white}{$\phantom{0}29$} & $\phantom{0}20$ & $\phantom{0}45$ & $1$ & $18$ & $\phantom{0}120$ \\
 & Lisa  & \cellcolor{white}{$\phantom{00}12$} & \cellcolor{white}{$\phantom{0}22$} & $\phantom{0}71$ & $\phantom{0}11$ &  & $\phantom{0}4$ & $\phantom{0}120$ \\
 & MoGuSer  & \cellcolor{white}{$\phantom{000}3$} & \cellcolor{white}{$\phantom{0}16$} & $\phantom{00}5$ & $\phantom{0}96$ &  &  & $\phantom{0}120$ \\
 & Cynthia  & \cellcolor{white}{$\phantom{00}14$} & \cellcolor{white}{$\phantom{0}26$} & $\phantom{0}35$ & $\phantom{0}45$ &  &  & $\phantom{0}120$ \\
 & Nike  & \cellcolor{white}{$\phantom{000}5$} & \cellcolor{white}{$\phantom{0}15$} & $\phantom{0}30$ & $\phantom{0}70$ &  &  & $\phantom{0}120$ \\
\midrule
pattern & ltlfsynt:full-N  & \cellcolor{yellow}{$\phantom{00}21$} & \cellcolor{yellow}{$\phantom{0}19$} &  &  &  &  & $\phantom{00}40$ \\
 & ltlfsynt:bfs-nopre  & \cellcolor{yellow}{$\phantom{00}21$} & \cellcolor{yellow}{$\phantom{0}19$} &  &  &  &  & $\phantom{00}40$ \\
 & ltlfsynt:bfs-os-nopre  & \cellcolor{yellow}{$\phantom{00}21$} & \cellcolor{yellow}{$\phantom{0}19$} &  &  &  &  & $\phantom{00}40$ \\
 & ltlfsynt:bfs  & \cellcolor{yellow}{$\phantom{00}21$} & \cellcolor{yellow}{$\phantom{0}19$} &  &  &  &  & $\phantom{00}40$ \\
 & ltlfsynt:bfs-os  & \cellcolor{yellow}{$\phantom{00}21$} & \cellcolor{yellow}{$\phantom{0}19$} &  &  &  &  & $\phantom{00}40$ \\
 & SyftMax  & \cellcolor{white}{$\phantom{00}20$} & \cellcolor{white}{$\phantom{0}17$} & $\phantom{00}1$ & $\phantom{00}2$ &  &  & $\phantom{00}40$ \\
 & Lydia  & \cellcolor{yellow}{$\phantom{00}21$} & \cellcolor{white}{$\phantom{0}17$} &  & $\phantom{00}2$ &  &  & $\phantom{00}40$ \\
 & LydiaSyft  & \cellcolor{yellow}{$\phantom{00}21$} & \cellcolor{yellow}{$\phantom{0}19$} &  &  &  &  & $\phantom{00}40$ \\
 & Lisa  & \cellcolor{white}{$\phantom{00}20$} & \cellcolor{white}{$\phantom{0}17$} & $\phantom{00}1$ & $\phantom{00}2$ &  &  & $\phantom{00}40$ \\
 & MoGuSer  & \cellcolor{yellow}{$\phantom{00}21$} & \cellcolor{yellow}{$\phantom{0}19$} &  &  &  &  & $\phantom{00}40$ \\
 & Cynthia  & \cellcolor{yellow}{$\phantom{00}21$} & \cellcolor{yellow}{$\phantom{0}19$} &  &  &  &  & $\phantom{00}40$ \\
 & Nike  & \cellcolor{yellow}{$\phantom{00}21$} & \cellcolor{yellow}{$\phantom{0}19$} &  &  &  &  & $\phantom{00}40$ \\
\midrule
random & ltlfsynt:full-N  & \cellcolor{yellow}{$\phantom{0}957$} & \cellcolor{yellow}{$278$} & $\phantom{00}1$ & $\phantom{00}1$ &  &  & $1237$ \\
 & ltlfsynt:bfs-nopre  & \cellcolor{white}{$\phantom{0}873$} & \cellcolor{yellow}{$278$} & $\phantom{0}52$ & $\phantom{0}34$ &  &  & $1237$ \\
 & ltlfsynt:bfs-os-nopre  & \cellcolor{white}{$\phantom{0}872$} & \cellcolor{yellow}{$278$} & $\phantom{0}58$ & $\phantom{0}29$ &  &  & $1237$ \\
 & ltlfsynt:bfs  & \cellcolor{yellow}{$\phantom{0}957$} & \cellcolor{yellow}{$278$} & $\phantom{00}1$ & $\phantom{00}1$ &  &  & $1237$ \\
 & ltlfsynt:bfs-os  & \cellcolor{yellow}{$\phantom{0}957$} & \cellcolor{yellow}{$278$} & $\phantom{00}1$ & $\phantom{00}1$ &  &  & $1237$ \\
 & SyftMax  & \cellcolor{white}{$\phantom{0}790$} & \cellcolor{white}{$259$} & $\phantom{0}15$ & $173$ &  &  & $1237$ \\
 & Lydia  & \cellcolor{white}{$\phantom{0}803$} & \cellcolor{white}{$266$} &  & $103$ &  & $65$ & $1237$ \\
 & LydiaSyft  & \cellcolor{white}{$\phantom{0}790$} & \cellcolor{white}{$273$} & $\phantom{0}12$ & $162$ &  &  & $1237$ \\
 & Lisa  & \cellcolor{white}{$\phantom{0}872$} & \cellcolor{white}{$265$} & $100$ &  &  &  & $1237$ \\
 & MoGuSer  & \cellcolor{white}{$\phantom{0}829$} & \cellcolor{white}{$257$} & $118$ & $\phantom{0}33$ &  &  & $1237$ \\
 & Cynthia  & \cellcolor{white}{$\phantom{0}192$} & \cellcolor{white}{$239$} & $247$ & $559$ &  &  & $1237$ \\
 & Nike  & \cellcolor{white}{$\phantom{0}898$} & \cellcolor{yellow}{$278$} & $\phantom{0}61$ &  &  &  & $1237$ \\
\midrule
all & ltlfsynt:full-N  & \cellcolor{white}{$1018$} & \cellcolor{white}{$342$} & $\phantom{0}20$ & $\phantom{0}17$ &  &  & $1397$ \\
 & ltlfsynt:bfs-nopre  & \cellcolor{white}{$\phantom{0}934$} & \cellcolor{white}{$344$} & $\phantom{0}56$ & $\phantom{0}63$ &  &  & $1397$ \\
 & ltlfsynt:bfs-os-nopre  & \cellcolor{white}{$\phantom{0}933$} & \cellcolor{yellow}{$354$} & $\phantom{0}61$ & $\phantom{0}49$ &  &  & $1397$ \\
 & ltlfsynt:bfs  & \cellcolor{yellow}{$1020$} & \cellcolor{white}{$344$} & $\phantom{00}5$ & $\phantom{0}28$ &  &  & $1397$ \\
 & ltlfsynt:bfs-os  & \cellcolor{yellow}{$1020$} & \cellcolor{yellow}{$354$} & $\phantom{00}4$ & $\phantom{0}19$ &  &  & $1397$ \\
 & SyftMax  & \cellcolor{white}{$\phantom{0}817$} & \cellcolor{white}{$305$} & $\phantom{0}39$ & $220$ & $1$ & $15$ & $1397$ \\
 & Lydia  & \cellcolor{white}{$\phantom{0}831$} & \cellcolor{white}{$311$} & $\phantom{0}24$ & $150$ & $1$ & $80$ & $1397$ \\
 & LydiaSyft  & \cellcolor{white}{$\phantom{0}818$} & \cellcolor{white}{$321$} & $\phantom{0}32$ & $207$ & $1$ & $18$ & $1397$ \\
 & Lisa  & \cellcolor{white}{$\phantom{0}904$} & \cellcolor{white}{$304$} & $172$ & $\phantom{0}13$ &  & $\phantom{0}4$ & $1397$ \\
 & MoGuSer  & \cellcolor{white}{$\phantom{0}853$} & \cellcolor{white}{$292$} & $123$ & $129$ &  &  & $1397$ \\
 & Cynthia  & \cellcolor{white}{$\phantom{0}227$} & \cellcolor{white}{$284$} & $282$ & $604$ &  &  & $1397$ \\
 & Nike  & \cellcolor{white}{$\phantom{0}924$} & \cellcolor{white}{$312$} & $\phantom{0}91$ & $\phantom{0}70$ &  &  & $1397$ \\
\bottomrule 
\end{tabular}

}
\end{table}

Figure~\ref{fig:Nike-vs-bfs-os} compare the best configuration of
\texttt{ltlfsynt} against Nike: there are no cases where Nike is
faster.  If we disable preprocessings and one-step (un)realizability
in \texttt{ltlfsynt}, the comparison is more balanced, as shown in
Figure~\ref{fig:Nike-vs-bfs-os-nopre}.  Note that we have kept
one-step (un)realizability enabled in this comparison, because Nike
uses it too.  This is the reason why \emph{pattern} benchmarks are
solved instantaneously by both tools.

\begin{figure}[tb]
  \includegraphics[width=\textwidth]{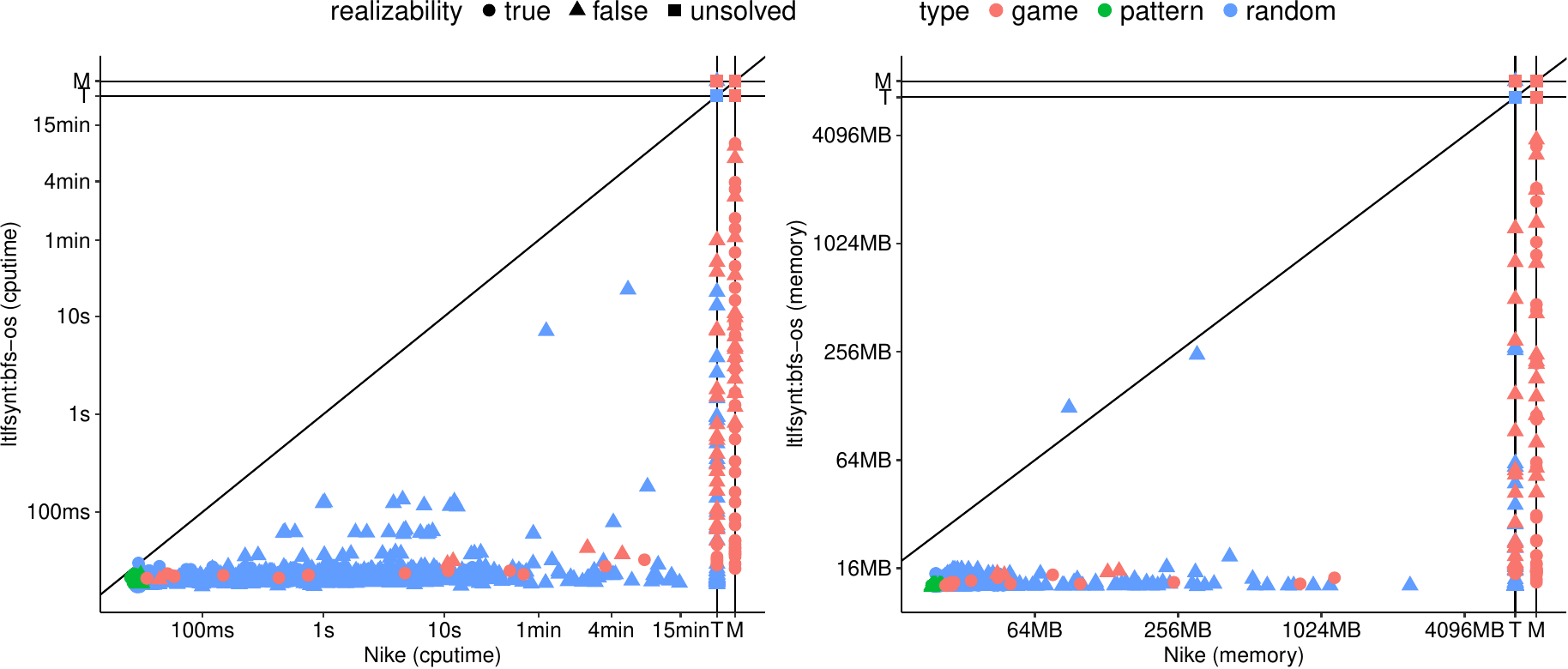}
  \caption{Scatter plots comparing time and memory usage of Nike against \texttt{ltlfsynt}'s best configuration.  Dots on the lines marked as \textsf{T} and \textsf{M} on the side represent timeouts or out-of-memory cases.\label{fig:Nike-vs-bfs-os}}
%\end{figure}
%\begin{figure}[tb]
  \includegraphics[width=\textwidth]{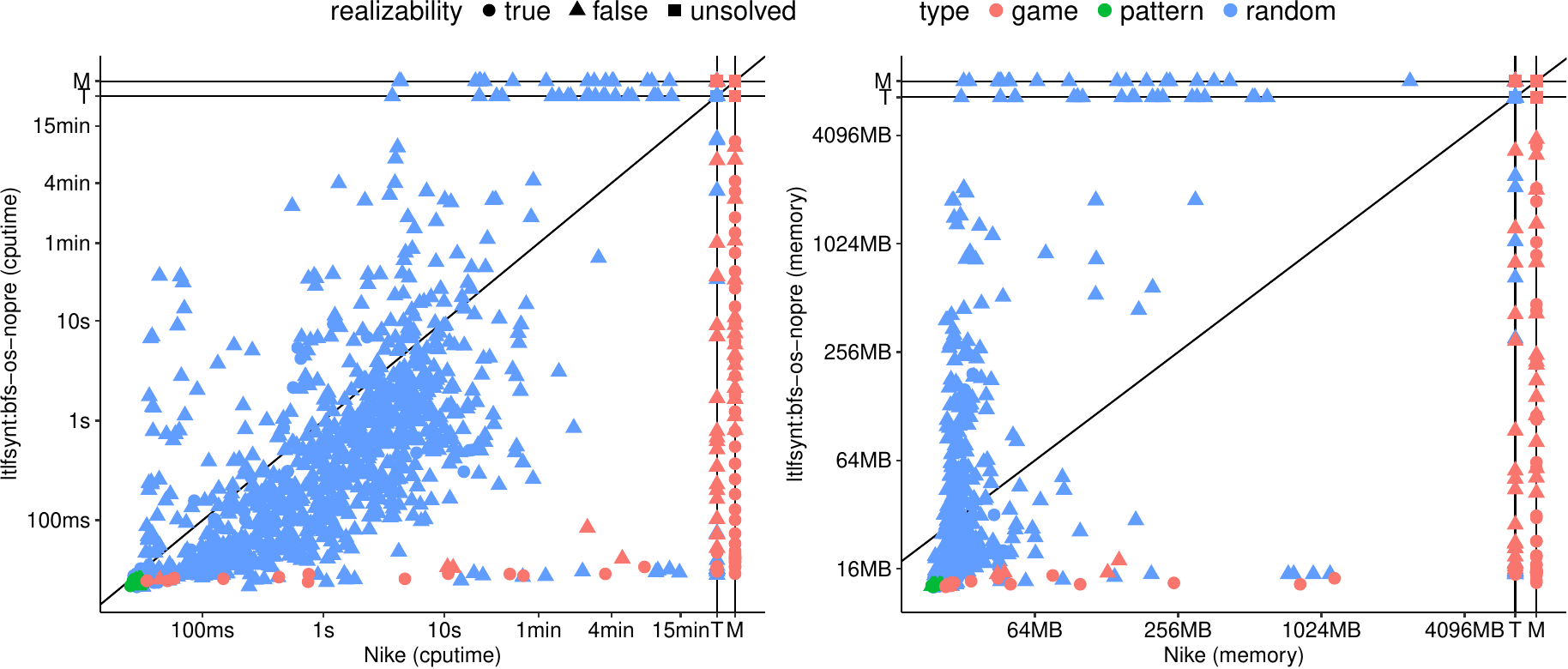}
  \caption{Scatter plots comparing time and memory usage of Nike
    against \texttt{ltlfsynt}'s on-the-fly construction but without
    preprocessing.\label{fig:Nike-vs-bfs-os-nopre}}
\end{figure}

Tables~\ref{tab:counter}, \ref{tab:counters}, and~\ref{tab:nim} look
at the runtime of the tools on \emph{game} benchmarks.  Values
highlighted in yellow are within 5\% of the minimum value of each
line.

Table~\ref{tab:counter} shows a family of specifications where
preprocessings are useless, and using one-step (un)realizability slows
things down.

Table~\ref{tab:counters} shows a family of specifications where
one-step (un)realizability is what allows \texttt{ltlfsynt} to solve
many more instance than other tools (even tools like Nike or LydiaSyft
who also implement one-step (un)realizability).  The suspicious
behavior of Lydia/LidyaSyft/SyftMax cycling between timeouts,
segmentation faults, and out-of-memory has been double-checked: this
is really how they terminated.

Finally, Table~\ref{tab:nim} shows very
impressive results by \texttt{ltlfsynt} on the challenging Nim family
of benchmarks: the highest configuration that third-party tools are
able to solve is \texttt{nim\textunderscore{}04\textunderscore{}01},
but \texttt{ltlfsynt} solves this instantaneously in all
configurations, and can handle much larger instances.

\begin{table}[tbp]
  \caption{Runtime of the different configurations on the Single Counter benchmark.\label{tab:counter}}
  \centering
  \adjustbox{width=\textwidth}{

\begin{tabular}[t]{>{}lrrrrrrrrrrrr}
\toprule
\multicolumn{1}{c}{} & \multicolumn{5}{c}{ltlfsynt} & \multicolumn{1}{c}{Lydia} & \multicolumn{1}{c}{SyftMax} & \multicolumn{1}{c}{LydiaSyft} & \multicolumn{1}{c}{Lisa} & \multicolumn{1}{c}{MoGuSer} & \multicolumn{1}{c}{Cynthia} & \multicolumn{1}{c}{Nike} \\
\cmidrule(l{3pt}r{3pt}){2-6} \cmidrule(l{3pt}r{3pt}){7-7} \cmidrule(l{3pt}r{3pt}){8-8} \cmidrule(l{3pt}r{3pt}){9-9} \cmidrule(l{3pt}r{3pt}){10-10} \cmidrule(l{3pt}r{3pt}){11-11} \cmidrule(l{3pt}r{3pt}){12-12} \cmidrule(l{3pt}r{3pt}){13-13}
file & full-N & bfs & bfs-os & bfs-os-nopre & bfs-nopre &   &   &   &   &   &   &  \\
\midrule
\cellcolor{white}{\textbf{counter\_01}} & \cellcolor{white}{\textcolor{black}{0.018}} & \cellcolor{white}{\textcolor{black}{0.020}} & \cellcolor{white}{\textcolor{black}{0.021}} & \cellcolor{white}{\textcolor{black}{0.025}} & \cellcolor{white}{\textcolor{black}{0.024}} & \cellcolor{white}{\textcolor{black}{0.044}} & \cellcolor{white}{\textcolor{black}{0.028}} & \cellcolor{white}{\textcolor{black}{0.131}} & \cellcolor{white}{\textcolor{black}{0.042}} & \cellcolor{yellow}{\textcolor{black}{0.016}} & \cellcolor{white}{\textcolor{black}{0.036}} & \cellcolor{white}{\textcolor{black}{0.035}}\\
\cellcolor{white}{\textbf{counter\_02}} & \cellcolor{yellow}{\textcolor{black}{0.018}} & \cellcolor{white}{\textcolor{black}{0.022}} & \cellcolor{white}{\textcolor{black}{0.022}} & \cellcolor{white}{\textcolor{black}{0.026}} & \cellcolor{white}{\textcolor{black}{0.025}} & \cellcolor{white}{\textcolor{black}{0.075}} & \cellcolor{white}{\textcolor{black}{0.033}} & \cellcolor{white}{\textcolor{black}{0.155}} & \cellcolor{white}{\textcolor{black}{0.125}} & \cellcolor{white}{\textcolor{black}{0.041}} & \cellcolor{white}{\textcolor{black}{0.498}} & \cellcolor{white}{\textcolor{black}{0.058}}\\
\bottomrule
\end{tabular}
}
\end{table}

\begin{table}[tbp]
  \caption{Runtime of the different configurations on the Double Counters benchmark.\label{tab:counters}}
  \centering
  \adjustbox{width=\textwidth}{

\begin{tabular}[t]{>{}lrrrrrrrrrrrr}
\toprule
\multicolumn{1}{c}{} & \multicolumn{5}{c}{ltlfsynt} & \multicolumn{1}{c}{Lydia} & \multicolumn{1}{c}{SyftMax} & \multicolumn{1}{c}{LydiaSyft} & \multicolumn{1}{c}{Lisa} & \multicolumn{1}{c}{MoGuSer} & \multicolumn{1}{c}{Cynthia} & \multicolumn{1}{c}{Nike} \\
\cmidrule(l{3pt}r{3pt}){2-6} \cmidrule(l{3pt}r{3pt}){7-7} \cmidrule(l{3pt}r{3pt}){8-8} \cmidrule(l{3pt}r{3pt}){9-9} \cmidrule(l{3pt}r{3pt}){10-10} \cmidrule(l{3pt}r{3pt}){11-11} \cmidrule(l{3pt}r{3pt}){12-12} \cmidrule(l{3pt}r{3pt}){13-13}
file & full-N & bfs & bfs-os & bfs-os-nopre & bfs-nopre &   &   &   &   &   &   &  \\
\midrule
\cellcolor{white}{\textbf{counters\_01}} & \cellcolor{yellow}{\textcolor{black}{0.017}} & \cellcolor{white}{\textcolor{black}{0.021}} & \cellcolor{white}{\textcolor{black}{0.021}} & \cellcolor{white}{\textcolor{black}{0.025}} & \cellcolor{white}{\textcolor{black}{0.025}} & \cellcolor{white}{\textcolor{black}{0.081}} & \cellcolor{white}{\textcolor{black}{0.036}} & \cellcolor{white}{\textcolor{black}{0.156}} & \cellcolor{white}{\textcolor{black}{0.156}} & \cellcolor{white}{\textcolor{black}{0.019}} & \cellcolor{white}{\textcolor{black}{0.065}} & \cellcolor{white}{\textcolor{black}{0.034}}\\
\cellcolor{white}{\textbf{counters\_02}} & \cellcolor{yellow}{\textcolor{black}{0.021}} & \cellcolor{yellow}{\textcolor{black}{0.022}} & \cellcolor{white}{\textcolor{black}{0.023}} & \cellcolor{white}{\textcolor{black}{0.025}} & \cellcolor{white}{\textcolor{black}{0.027}} & \cellcolor{white}{\textcolor{black}{0.221}} & \cellcolor{white}{\textcolor{black}{0.063}} & \cellcolor{white}{\textcolor{black}{0.218}} & \cellcolor{white}{\textcolor{black}{0.077}} & \cellcolor{white}{\textcolor{black}{0.040}} & \cellcolor{white}{\textcolor{black}{0.224}} & \cellcolor{white}{\textcolor{black}{0.051}}\\
\bottomrule
\end{tabular}
}
\end{table}

\begin{table}[tbp]
  \caption{Runtime of the different configurations on the Nim benchmark.\label{tab:nim}}
  \centering
  \adjustbox{totalheight=\textheight-2\baselineskip}{

\begin{tabular}[t]{>{}lrrrrrrrrrrrr}
\toprule
\multicolumn{1}{c}{} & \multicolumn{5}{c}{ltlfsynt} & \multicolumn{1}{c}{Lydia} & \multicolumn{1}{c}{SyftMax} & \multicolumn{1}{c}{LydiaSyft} & \multicolumn{1}{c}{Lisa} & \multicolumn{1}{c}{MoGuSer} & \multicolumn{1}{c}{Cynthia} & \multicolumn{1}{c}{Nike} \\
\cmidrule(l{3pt}r{3pt}){2-6} \cmidrule(l{3pt}r{3pt}){7-7} \cmidrule(l{3pt}r{3pt}){8-8} \cmidrule(l{3pt}r{3pt}){9-9} \cmidrule(l{3pt}r{3pt}){10-10} \cmidrule(l{3pt}r{3pt}){11-11} \cmidrule(l{3pt}r{3pt}){12-12} \cmidrule(l{3pt}r{3pt}){13-13}
file & full-N & bfs & bfs-os & bfs-os-nopre & bfs-nopre &   &   &   &   &   &   &  \\
\midrule
\cellcolor{white}{\textbf{nim\_01\_01}} & \cellcolor{yellow}{\textcolor{black}{0.017}} & \cellcolor{white}{\textcolor{black}{0.021}} & \cellcolor{white}{\textcolor{black}{0.020}} & \cellcolor{white}{\textcolor{black}{0.025}} & \cellcolor{white}{\textcolor{black}{0.025}} & \cellcolor{white}{\textcolor{black}{0.090}} & \cellcolor{white}{\textcolor{black}{0.036}} & \cellcolor{white}{\textcolor{black}{0.205}} & \cellcolor{white}{\textcolor{black}{0.037}} & \cellcolor{white}{\textcolor{black}{0.036}} & \cellcolor{white}{\textcolor{black}{0.026}} & \cellcolor{white}{\textcolor{black}{0.044}}\\
\cellcolor{white}{\textbf{nim\_01\_02}} & \cellcolor{yellow}{\textcolor{black}{0.017}} & \cellcolor{white}{\textcolor{black}{0.022}} & \cellcolor{white}{\textcolor{black}{0.022}} & \cellcolor{white}{\textcolor{black}{0.026}} & \cellcolor{white}{\textcolor{black}{0.026}} & \cellcolor{white}{\textcolor{black}{0.163}} & \cellcolor{white}{\textcolor{black}{0.047}} & \cellcolor{white}{\textcolor{black}{0.249}} & \cellcolor{white}{\textcolor{black}{0.058}} & \cellcolor{white}{\textcolor{black}{0.253}} & \cellcolor{white}{\textcolor{black}{0.072}} & \cellcolor{white}{\textcolor{black}{0.058}}\\
\bottomrule
\end{tabular}
}
\end{table}

A more detailed analysis of the benchmark results can be found in
directory \texttt{ltlfsynt-analysis/} of the associated
artifact~\cite{duret.25.zenodo}.

\fi
\end{document}